\documentstyle[aps]{revtex}
\begin{document}
\title{Strange meson-nucleon states in the quark potential model }
\author{Hai-Jun Wang and Jun-Chen Su}
\address{Center for Theoretical Physics, School of Physics, Jilin University,\\
Changchun 130023, China}
\maketitle

\begin{abstract}
The quark potential model and resonating group method are used to
investigate the $\overline{K}N$ bound states and/or resonances. The model
potential consists of the t-channel and s-channel one-gluon exchange
potentials and the confining potential with incorporating the QCD
renormalization correction and the spin-orbital suppression effect in it. It
was shown in our previous work that by considering the color octet
contribution, use of this model to investigate the $KN$ low energy elastic
scattering leads to the results which are in pretty good agreement with the
experimental data. In this paper, the same model and method are employed to
calculate the masses of the $\overline{K}N$ bound systems. For this purpose,
the resonating group equation is transformed into a standard Schr\"odinger
equation in which a nonlocal effective $\overline{K}N$ interaction potential
is included. Solving the Schr\"odinger equation by the variational method,
we are able to reproduce the masses of some currently concerned $\overline{K}%
N$ states and get a view that these states possibly exist as $\overline{K}N$
molecular states. For the $KN$ system, the same calculation gives no support
to the existence of the resonance $\Theta ^{+}(1540)$ which was announced
recently.
\end{abstract}

\section{Introduction}

Recently, the $\overline{K}\ N$ systems has attracted much interest in
investigations of the puzzle of exotic baryon state $\Lambda (1405)$ [1-9].
The puzzle of $\Lambda (1405)$ came from the obvious discrepancy between the
downward shift of the $1S$ level of the kaonic hydrogen atom which was
determined from the measurement of atomic x rays [10-12] and the upward
shift measured from low energy $\overline{K}N$ scattering [13]. Although the
discrepancy itself has been resolved recently by the elaborate measurement
of x rays from the atom with an upward shift of the $1S$ level [14,15], how
to understand the nature of the state has not reached its last word yet. The
state seems to be able to be interpreted as an elementary baryon, i.e., a
three quark state belongs to the $70^{-}$ multiplet, a meson-baryon
composite being a $\overline{K}N$ bound state or a $\pi \Sigma $ resonance,
or a five quark bound state. In Ref. [16], the authors investigated the
state $\Lambda (1405)$ by employing a coupled-channel potential model with
introducing a separable Yukawa-type meson-baryon potential. From fitting to
low-energy $\overline{K}N$ scattering data, they obtained two sets of
parameters: one allows them to interpret the state $\Lambda (1405)$ as the $%
70^{-}$ three-quark state strongly coupled with $\overline{K}N$ and $\pi
\Sigma $; another to interpret the state as a $\pi \Sigma $ resonance and/or
an unstable $\overline{K}N$ bound state. In Ref. [17], the $s$-wave
meson-baryon interaction in the strangeness $S=-1$ sector was studied by
means of a coupled channel method by using the lowest chiral Lagrangian. By
a good fit to the scattering data, the authors conclude that the $\Lambda
(1405)$ and $\Lambda (1670)$ may be identified with $\overline{K}N$ and $%
K\Xi $ quasibound states, respectively. The resonances $\Lambda (1405)$ and $%
\Lambda (1670)$ were also investigated in Ref. [18] in the Bethe-Salpeter
coupled-channel formalism by utilizing the interaction given by the lowest
perturbative chiral Lagrangian. In addition, the resonances $\Lambda (1405)$
and $\Sigma (1620)$ were ever studied in the QCD sum rule approach by using
multiquark interpolating fields $((q\overline{q})(qqq))$ [19]. In this
study, the above resonances are considered as multiquark states. In Ref.
[20], the two-pole structure of $\Lambda (1405)$ was studied from the
reaction $K^{-}p\rightarrow \pi ^0\pi ^0\Sigma ^0$ in the energy region of $%
p_{K^{-}}=514$ to $750$ $MeV/c$ by using the chiral unitary theory. The
possible existence of deeply bound kaonic states was recently investigated
for few-body systems by assuming a kaon-nucleon interaction of Gaussian type
with a fixed width [21]. In this investigation, it was concluded that the $%
\Lambda (1405)$\ may be viewed as a $\overline{K}N$\ bound state. However, a
critical analysis given in Ref. [22] on the KEK and FINUDA experiments [23]
indicates that the experimental confirmation of the existence of deeply
bound kaonic states is still under debate.

In our previous work [24], we have elaborately calculated the $KN$ and $%
\overline{K}N$ elastic scattering phase shifts. The results are in quite
good agreement with the experimental data and show that for the $\overline{K}%
N$ system, the $s$-wave shifts exhibit strong attractive interactions in the 
$s$-channel scattering, while, for the $KN$ system, only the $P_{01}$ and $%
D_{03}$ states have small positive phase shifts which imply that the
attractive interactions in these states are weak. The aim of this paper is
to investigate the $\overline{K}N$ bound states and resonances by using the
quark potential model and resonating group method (RGM) [25], and to discuss
whether the pentaquark $\Theta ^{+}(1540)$ [26-29] is possible to exist or
not. The potential used is composed of the $t$-channel one gluon exchange
potential (OGEP) [30] and the $s-$channel OGEP [24, 31-33] as well as a
phenomenological confining potential. The two OGEPs were derived from QCD in
the nonrelativistic approximation of order $p^2/m^2$ and contain
spin-independent terms such as the Coulomb, velocity-dependent terms and
spin-dependent terms such as the spin-spin interaction, spin-orbital
coupling and tensor force terms. All these terms are taken into account in
our investigation as should be done in a theoretically consistent treatment.
Especially, the $s$-channel OGEP is necessary to be considered for the $%
\overline{K}N$ interaction. With this potential and considering the
contribution of color octet of clusters $K(\overline{K})$ and $N$, it was
demonstrated in our previous works [24, 32, 33] that the model is not only
able to quite well reproduce the $KN$ low-energy elastic scattering data,
but also to give reasonable results for the $\pi \pi $ low-energy scattering
and $K\overline{K\text{ }}$ bound states. The results obtained in our
previous works encourage us to study $\overline{K}N$ and $KN$ resonances $%
\Lambda (1405),$ $\Lambda (1600)$, $\Lambda (1670)$, $\Sigma (1385)$, $%
\Sigma (1620)$ and $\Theta ^{+}(1540)$ in a consistent way within the
framework of the aforementioned potential model and RGM.

\section{Formalism}

In this section, we briefly describe the potential model and resonating
group method (RGM). According to the quark model, the $KN$ ( $\overline{K}N$
) system may be treated as two quark clusters: the $K$-cluster $(q\overline{s%
})$ ( the $\overline{K}$-cluster $(\overline{q}s)$) and the $N$-cluster ($%
qqq)$ where $q=u$ or $d$. The effective $KN$ $(\overline{K}N)$ interaction
potential may be extracted from the following Schr\"odinger equation for the
interacting $q^4\overline{s}$ $(q^3\overline{q}s)$ system by the RGM [25] 
\begin{equation}
(T+V)\Psi =E\Psi  \eqnum{1}
\end{equation}
where $E$, $T$, $V$ and $\Psi $ stand for the total energy, the kinetic
energy, the interaction potential and the wave function of the multiquark
system, respectively.

\subsection{Interquark potential}

In the center of mass frame, the kinetic energy is given by 
\begin{equation}
T=\sum_{i=1}^5\frac{\vec p_i^2}{2m_i}-T_c  \eqnum{2}
\end{equation}
where $T_c$ represents the center of mass kinetic energy, The interaction
potential is assumed to be 
\begin{equation}
V=\sum_{i<j=1}^5(V_{ij}^t+V_{ij}^s+V_{ij}^c)  \eqnum{3}
\end{equation}
where $V_{ij}^t$, $V_{ij}^s$ and $V_{ij}^c$ denote the $t$-channel OGEP, the 
$s$-channel OGEP and the confining potentials, respectively. They are
separately described in the following.

The $t$-channel OGEP represented in the momentum space is [24, 30] 
\begin{equation}
\begin{tabular}{l}
$V_{ij}^t=\frac{4\pi \alpha _sC_{ij}^t}{(\vec q-\vec k)^2}\{1-\frac{\vec P^2%
}{m_{ij}^2}-\frac{(m_i^2+m_j^2)}{8m_i^2m_j^2}(\vec q-\vec k)^2+\frac{\left(
m_i-m_j\right) }{2m_im_jm_{ij}}\vec P\cdot (\vec q+\vec k)$ \\ 
$+\frac{(\vec q+\vec k)^2}{4m_im_j}+\frac i{4m_{ij}}[\vec P\cdot (\vec q-%
\vec k)\cdot (\frac{\vec \sigma _i}{mi}-\frac{\vec \sigma _j}{m_j})]-\frac{(%
\vec q-\vec k)^2}{4m_im_j}\vec \sigma _i\cdot \vec \sigma _j$ \\ 
$+\frac i{4m_{ij}}(\vec q\times \vec k)\cdot [(2+\frac{m_j}{m_i})\vec \sigma
_i+(2+\frac{m_i}{m_j})\vec \sigma _j]+\frac{(\vec q-\vec k)\cdot \vec \sigma
_i(\vec q-\vec k)\cdot \vec \sigma _j}{4m_im_j}\}$%
\end{tabular}
\eqnum{4}
\end{equation}
where $m_{ij}=m_i+m_j$ with $m_i$ and $m_j$ being the masses of $i$-th and $%
j $-th quarks respectively, $\alpha _s$ is the QCD fine structure constant, $%
\vec \sigma _i$and $\vec \sigma _j$are the spin Pauli matrices for $i$-th
and $j$-th quarks, $C_{ij}^t$ is the $t$-channel color matrix defined as

\begin{equation}
C_{ij}^t=\{ 
\begin{tabular}{ll}
$\frac{\lambda _i^a}2\frac{\lambda _j^a}2(\frac{\lambda _i^{a*}}2\frac{%
\lambda _j^{a*}}2),$ & $\text{for }qq(\overline{q}\overline{q})$ \\ 
$-\frac{\lambda _i^a}2\frac{\lambda _j^{a^{*}}}2,$ & $\text{for }q\overline{q%
}$%
\end{tabular}
\eqnum{5}
\end{equation}
with $\lambda ^a$ being the Gell-Mann matrix, $\vec P,\vec k$ and $\vec q$
are respectively the total momentum, the initial state relative momentum and
the final state relative momentum of the two interacting particles.

The $s$-channel OGEP is [24, 31-33] 
\begin{equation}
\begin{tabular}{l}
$V_{ij}^s=\frac{\pi \alpha _sF_{ij}^sC_{ij}^s}{2mm^{^{\prime }}}[(3+\vec 
\sigma _i\cdot \vec \sigma _j)-\frac{5(m^2+m^{\prime 2})-4mm^{\prime }}{%
8m^2m^{\prime 2}}\vec P^2-\frac{2\vec k^2}{m^2}-\frac{2\vec q^2}{m^{\prime 2}%
}$ \\ 
$-(\frac{(m^2+m^{\prime 2})}{8m^2m^{\prime 2}}+\frac{\vec k^2}{m^2}+\frac{%
\vec q^2}{m^{\prime 2}})\vec \sigma _i\cdot \vec \sigma _j+\frac i{%
4m^{\prime 2}}(\vec P\times \vec q)\cdot (\vec \sigma _i-\vec \sigma _j)$ \\ 
$-\frac i{4m^2}(\vec P\times \vec k)\cdot (\vec \sigma _i-\vec \sigma _j)-%
\frac{\left( m-m^{\prime }\right) ^2}{4m^2}\vec P\cdot \vec \sigma _i\vec P%
\cdot \vec \sigma _j$ \\ 
$+\frac 1{4m^2}(\vec P\cdot \vec \sigma _i\vec k\cdot \vec \sigma _j-\vec k%
\cdot \vec \sigma _i\vec P\cdot \vec \sigma _j+4\vec k\cdot \vec \sigma _i%
\vec k\cdot \vec \sigma _j)$ \\ 
$+\frac 1{4m^{\prime 2}}(\vec P\cdot \vec \sigma _i\vec q\cdot \vec \sigma
_j-\vec q\cdot \vec \sigma _i\vec P\cdot \vec \sigma _j+4\vec q\cdot \vec 
\sigma _i\vec q\cdot \vec \sigma _j)]$%
\end{tabular}
\eqnum{6}
\end{equation}
where $m$ and $m^{\prime }$ denote the quark (antiquark) masses before and
after annihilation respectively, $C_{ij}^s$ and $F_{ij}^s$ are respectively
the $s$-channel color and flavor matrices defined by

\begin{equation}
C_{ij}^s=\frac 1{24}(\lambda _i^a-\lambda _j^{a^{*}})^2  \eqnum{7}
\end{equation}
here $\lambda _i^a$ are the Gell-Mann matrices for $i$-th quark and

\begin{equation}
F_{ij}^s=\frac 13-(\frac 12\vec \tau _i\cdot \vec \tau
_j+V_i^{-}V_j^{+}+V_i^{+}V_j^{-}+U_i^{-}U_j^{+}+U_i^{+}U_j^{-}+\frac 32%
Y_iY_j)  \eqnum{8}
\end{equation}
here $\stackrel{\rightharpoonup }{\tau _i}$ are the isospin Pauli matrices
for $i$-th particle, $Y_i$ the hypercharge operators, $V_i^{+}$ and $V_i^{-}$%
( $U_i^{+}$ and $U_i^{-}$) represent the rising and lowering operators of
the $V$-spin ($U$-spin) respectively. The $s$-channel OGEP gives
nonvanishing $S$-matrix elements only for $\overline{K}N$ system

The confining potential, as was done in Ref. [32], is taken to be a harmonic
oscillator one. In momentum space, it is represented as

\begin{equation}
V_{ij}^c=C_{ij}^t(2\pi )^3\mu _{ij}\omega ^2\nabla _k^2\delta ^3(\vec q-\vec 
k)  \eqnum{9}
\end{equation}
where $\mu _{ij}$ is the reduced mass of two interacting particles and $%
\omega $ the force-strength parameter.

\subsection{Resonating group method}

In application of resonating group method to investigate the system of
composite particles, we need to give the basis wave function of the system.
Let us first construct the basis function of $KN$ system from the wave
functions of clusters $(q\overline{s})$ and ($qqq)$. Since there are
identical particles between the two clusters, the basis function of the
system may be represented as 
\begin{equation}
\begin{tabular}{l}
$\Phi _{TM\frac 12m}(\vec p_1,\vec p_2,\vec p_3,\vec p_4,\vec p_5;\vec \rho
) $ \\ 
$={\cal A}\Psi _{TM\frac 12m}(1,2,3,4,5)R(\vec p_1,\vec p_2,\vec p_3,\vec p%
_4,\vec p_5;\vec \rho )$%
\end{tabular}
\eqnum{10}
\end{equation}
where the three quarks in the $N$-cluster are labeled as $1,2,3$ and the
quark and antiquark in the $K$-cluster (or the antiquark and quark in the $%
\overline{K}$-cluster) as $4$ and $5$, $\Psi _{TM\frac 12m}(1,2,3,4,5)$ and $%
R(\vec p_1,\vec p_2,\vec p_3,\vec p_4,\vec p_5;\vec \rho )$ represent the
color-isospin-spin wave function and the position space wave function
respectively which are constructed from the color-isospin-spin wave
functions and the coordinate space wave functions of nucleon and kaon, 
\begin{equation}
{\cal A=}\frac 1{\sqrt{4}}(1-P_{14}-P_{24}-P_{34})  \eqnum{11}
\end{equation}
denotes the antisymmetrized operator in which $P_{j4}$ ($j=1,2,3$) symbolize
the interchange operators. For the $\overline{K}N$ system, noticing that
there is no identical particles between the two clusters $(\overline{q}s)$
and $(qqq)$, the basis wave function of the system may simply be written out
from Eq. (10) by setting ${\cal A}=1.$ In this case, $\Psi _{TM\frac 12%
m}(1,2,3,4,5)$ and $R(\vec p_1,\vec p_2,\vec p_3,\vec p_4,\vec p_5;\vec \rho
)$ are the color-isospin-spin wave function and the position space wave
function constructed from the corresponding wave functions of nucleon and
antikaon.

When kaon (antikaon) and nucleon interact, the color states of the two quark
clusters are possibly polarized. In this case, the kaon (antikaon) and
nucleon should not be viewed as pure color singlet objects even though the
whole system is kept in color singlet and therefore the interaction taking
place in the color octet must be taken into account. Consideration of color
octet channel interactions has been proved to be important to reproduce
experimental data in the investigations of hadron productions and decays
[34-36]. Thus, the color-spin-isospin wave function $\Psi _{TM\frac 12%
m}(1,2,3,4,5)$ of the whole system may be given by the color singlet part $%
\Psi _{TM\frac 12m}^{(1)}(1,2,3,4,5)$ or the color octet part $\Psi _{TM%
\frac 12m}^{(2)}(1,2,3,4,5)$ formed by the color singlets or color octets of
the two clusters. In principle, we may test a general color structure of
system under consideration which is given by the following linear
combination 
\begin{equation}
\Psi _{TM\frac 12m}(1,2,3,4,5)=\alpha \Psi _{TM\frac 12m}^{(1)}(1,2,3,4,5)+%
\beta \Psi _{TM\frac 12m}^{(2)}(1,2,3,4,5)  \eqnum{12}
\end{equation}
where the coefficients $\alpha $ and $\beta $ are required to satisfy 
\begin{equation}
\left| \alpha \right| ^2+\left| \beta \right| ^2=1.  \eqnum{13}
\end{equation}
The wave functions $\Psi _{TM\frac 12m}^{(1)}(1,2,3,4,5)$ and $\Psi _{TM%
\frac 12m}^{(2)}(1,2,3,4,5)$, as described in Appendix A, were constructed
in Ref. [24] by the antisymmetric requirement for the wave functions of
identical particles in nucleon.

Because we limit our discussion to the interaction in the low-energy regime,
it is appropriate to write the position space basis function of the $KN$ or $%
\overline{K}N$ system in the form 
\begin{equation}
R(\vec p_1,\vec p_2,\vec p_3,\vec p_4,\vec p_5;\vec \rho )=\phi
_{os}^{\left( +\right) }(\vec p_1,\vec \rho )\phi _{os}^{\left( +\right) }(%
\vec p_2,\vec \rho )\phi _{os}^{\left( +\right) }(\vec p_3,\vec \rho )\phi
_{os}^{\left( -\right) }(\vec p_4,\vec \rho )\phi _{os}^{\left( -\right) }(%
\vec p_5,\vec \rho )  \eqnum{14}
\end{equation}
where $\phi _{os}^{\left( +\right) }(\vec p_i,\vec \rho )$ and $\phi
_{os}^{\left( -\right) }(\vec p_j,\vec \rho )$ are the lowest-lying harmonic
oscillator states of the $N$-cluster and $K(\overline{K})$-cluster given in
the momentum space, 
\begin{equation}
\phi _{os}^{\left( \pm \right) }(\vec p_i,\vec \rho )=(2\sqrt{\pi }%
b_i)^{3/2}\exp (-\frac 12b_i^2\vec p_i^2\mp i\lambda _{\pm }\vec p_i\cdot 
\vec \rho )  \eqnum{15}
\end{equation}
in which $\vec \rho $ is the vector representing the separation between the
centers of mass of the two clusters and parameters $\lambda _{\pm }$ are
defined by

\begin{equation}
\lambda _{-}=\beta _1=\frac{3m_1}{4m_1+m_2},\lambda _{+}=\beta _2=\frac{%
m_1+m_2}{4m_1+m_2}  \eqnum{16}
\end{equation}
here $m_1$ denotes the mass of $d$ or $u$ quark, $m_2$ the mass of strange
quark. The wave function in Eq. (14) can be represented through the cluster
coordinates in the form 
\begin{equation}
R(\vec p_1,\vec p_2,\vec p_3,\vec p_4,\vec p_5;\vec \rho )=X_K(\vec q)X_N(%
\vec k_1,\vec k_2)\Gamma (\vec Q,\vec \rho )Z_{CM}(\vec P)  \eqnum{17}
\end{equation}
where $X_K(\vec q)$ and $X_N(\vec k_1,\vec k_2)$ are the internal motion
wave functions of the $K(\overline{K})$-cluster $(q\overline{s})$ ($(%
\overline{q}s)$) and the $N$-cluster $(qqq)$ with $\vec q$ and $\vec k_1,%
\vec k_2$ being the relative momenta in the clusters $(q\overline{s})$ ($(%
\overline{q}s)$) and ($qqq)$ respectively, $\Gamma (\vec Q,\vec \rho )$ is
the wave function describing the relative motion between the two clusters
with $\stackrel{\rightharpoonup }{Q}$ being the relative momentum of the two
clusters and $Z_{CM}(\vec P)$ the wave function for the center-of-mass
motion of the whole system in which $\vec P$ is the total momentum of the
system. According to the RGM, the wave function of the two clusters may be
represented in the form

\begin{equation}
\overline{\Psi }_{TMsm}=\int d^3\rho \Phi _{TM\frac 12m_s}(\vec p_1,\vec p_2,%
\vec p_3,\vec p_4,\vec p_5;\vec \rho )f(\vec \rho )  \eqnum{18}
\end{equation}
where $\Phi _{TM\frac 12m_s}(\vec p_1,\vec p_2,\vec p_3,\vec p_4,\vec p_5;%
\vec \rho )$ is the basis function defined in Eq. (10) and $f(\vec \rho )$
is the unknown function describing the relative motion of the two clusters.
On substituting the above wave function in Eq. (1), according to the
well-known procedure, one may derive a resonating group equation satisfied
by the function $f(\vec \rho )$ which is obtained from Eq. (1) by
subtracting the internal motion of the two clusters, 
\begin{equation}
\int d^3\rho [T_r(\vec \rho ,\vec \rho ^{\prime })+V_r(\vec \rho ,\vec \rho
^{\prime })-E_rN_r(\vec \rho ,\vec \rho ^{\prime })]f(\vec \rho ^{\prime })=0
\eqnum{19}
\end{equation}
where 
\begin{equation}
X(\vec \rho ,\vec \rho ^{\prime })=\int \prod_{i=1}^5\frac{d^3p_k}{(2\pi )^3}%
\frac{d^3p_k^{\prime }}{(2\pi )^3}\langle \Phi _{TM\frac 12m}(\vec p_1,\vec p%
_2,\vec p_3,\vec p_4,\vec p_5;\vec \rho )\mid {\cal A}X\mid \Phi _{TM\frac 12%
m}(\vec p_1,\vec p_2,\vec p_3,\vec p_4,\vec p_5;\vec \rho ^{\prime })\rangle
\eqnum{20}
\end{equation}
here $X$ stands for $T,V$ or $I$ (the unity). The expressions of the
functions $T_r(\vec \rho ,\vec \rho ^{\prime })$, $V_r(\vec \rho ,\vec \rho
^{\prime })$ and $N_r(\vec \rho ,\vec \rho ^{\prime })$ are not shown here
but will be used to derive the effective potential as described soon later.

\subsection{Schr\"odinger equation and effective $KN(\overline{K}N)$
potential}

The resonating group equation in Eq. (19) is not of the standard form of
Schr\"odinger equation since the normalization function $N_r(\vec \rho ,\vec 
\rho ^{\prime })$ is not unity. It can be converted into a Schr\"odinger
equation by the following transformation 
\begin{equation}
f(\vec \rho )=\int d^3R\Gamma (\vec \rho ,\vec R)\overline{\Psi }(\vec R) 
\eqnum{21}
\end{equation}
where 
\begin{equation}
\Gamma (\vec \rho .\vec R)=\frac 1{\sqrt{2}(2\pi )^3}(\frac{3\beta _2}{\pi
b^2})^{3/4}\int d^3ke^{\frac 1{6\beta _2}b^2\vec k^2+i\vec k\cdot (\vec \rho
-\vec R)}  \eqnum{22}
\end{equation}
in which $b$ is the harmonic oscillator size parameter and $\overline{\Psi }(%
\vec R)$ will be identified with the wave function describing the relative
motion of the two clusters. On inserting Eq. (21) into Eq. (19), it is not
difficult to get 
\begin{equation}
-\frac 1{2\mu }\nabla _{\vec R}^2\overline{\Psi }(\vec R)+\int
d^3R^{^{\prime }}V(\vec R,\vec R^{\prime })\overline{\Psi }(\vec R^{\prime
})=\varepsilon \overline{\Psi }(\vec R)  \eqnum{23}
\end{equation}
where $\varepsilon $ stands for the binding energy, $\mu $ denotes the
reduced mass of nucleon and kaon (antikaon) and 
\begin{equation}
V(\vec R,\vec R^{\prime })=V^t(\vec R,\vec R^{\prime })+V^s(\vec R,\vec R%
^{\prime })+V^c(\vec R,\vec R^{\prime })  \eqnum{24}
\end{equation}
is the nonlocal $KN$ ( $\overline{K}N$ ) effective interaction potential in
which $V^t(\vec R,\vec R^{\prime })$, $V^s(\vec R,\vec R^{\prime })$ and $%
V^c(\vec R,\vec R^{\prime })$ are generated by the $t$-channel OGEP, the $s$%
-channel OGEP and the confining potential respectively. The potential $V(%
\vec R,\vec R^{\prime })$ in the Schr\"odinger equation is connected with
the potential $V(\vec \rho ,\vec \rho ^{\prime })$ appearing in the
resonating group equation by the following formula. 
\begin{equation}
V(\vec R,\vec R^{\prime })=\int d^3\rho d^3\rho ^{^{\prime }}\Gamma (\vec R,%
\vec \rho )V(\vec \rho ,\vec \rho ^{\prime })\Gamma (\vec \rho ^{\prime },%
\vec R^{\prime }).  \eqnum{25}
\end{equation}
Through a lengthy derivation, we obtain an explicit expression of the
potential $V(\vec R,\vec R^{\prime })$ which is displayed in Appendix B.

\section{Numerical results}

As mentioned before, use of the above effective potential to calculate the $%
KN$ elastic scattering phase shifts leads to the results which are in
reasonable agreement with experimental data. In this paper, we recalculate
the phase shifts and find that a quite good fit to the experimental data can
also be achieved by appropriately increasing both the coupling constant and
the size parameter of harmonic oscillator with remaining other parameters
unchanged. Some phase shifts selected here are shown in Figs.1-4 in which
Figs. 1 and 2 give the theoretical prediction for some $\overline{K}N$ low
energy elastic scattering phase shifts. These phase shifts are all positive,
implying that the interactions in the relevant states are attractive and
therefore possibly form bound states or resonances in those states. It is
noted that in our present investigation, besides the color octet mechanism,
the QCD renormalization effect and the spin-orbital suppression are also
taken into account. In the calculations, the theoretical parameters are
taken to be: the coupling constant $\alpha _s^0=0.527$, the force strength
of confinement $\omega =0.2$ $GeV$, the harmonic oscillator size parameter $%
b=0.483$, the constituent quark masses $m_u=m_d=350MeV$ and $m_s=550$ $MeV$,
the color combination coefficient $\alpha =0.915$, the scale parameter of
QCD renormalization $\mu =0.195$ $GeV$ and the parameter of spin-orbital
suppression $\gamma =0.45$. Except for the parameter $\alpha ,$ all the
parameters are consistently used in calculations of $S=-1$ $\overline{K}N$
states and $S=+1$ $KN$ states.

The mass spectrum of $\overline{K}N$ bound and resonant states are
calculated from the Schr\"odinger equation written in Eq. (23) by the
variational method. The trial wave functions are chosen to be

\begin{equation}
\Psi (J,T)=\frac 1{\sqrt{2J+1}}\sum\limits_{nLS}\sum%
\limits_{m_lm}u(n,L,S,)C_{Lm_lSm}^{Jm_J}R_{nLm_l}\Psi _{TSm}  \eqnum{26}
\end{equation}
where $R_{nLm_l}$ stands for the coordinate space basis functions which are,
as usual, taken to be the harmonic oscillator wave functions, $\Psi _{TSm}$
represents the spin-isospin wave function of the two clusters with $S=1/2$,
and $u(n,L,S)$ are the unknown combination coefficients which can be
determined by solving the Schr\"odinger equation. Here the quantum numbers
of a state is represented by the conventional notation $L_{T\ 2J}(J^P)$
where $L$,$\ J,T$ and $P$ designate the orbital angular momentum, the total
angular momentum, the isospin and parity of the state. It should be noted
that the spin-isospin wave function $\Psi _{TSm}$ in Eq. (26) is written
formally. In the practical calculation, the effective potential in Eq. (23)
is given by the matrix element of the potential operator shown in Appendix B
between the color-spin-isospin wave function represented in Eq. (12) and
Appendix A. In this kind of calculation, the wave function $\Psi _{TSm}$ in
Eq. (26) is replaced by the color-spin-isospin wave function mentioned
above. Another point we would like to note is that in solving the
Schr\"odinger equation, the series over $n$ in Eq. (26) is cut off by a
limitation of $N=2n+L$. We found that to obtain sufficiently accurate
results, the $N$ is unnecessarily taken to be large. To solve the binding
energy from the Schr\"odinger equation, we choose the dimensionless
parameter $a=\sqrt{m\omega }$ as the variational parameter which is
determined by the following stationary condition

\begin{equation}
\frac{\partial \varepsilon }{\partial a}=0.  \eqnum{27}
\end{equation}
The mass of a $\overline{K}$ $(K)N$ bound state or resonance is given by 
\begin{equation}
M=m_1+m_2+\varepsilon  \eqnum{28}
\end{equation}
where $m_1$ and $m_2$ are the masses of constituent $K(\overline{K})$ and $N$
clusters.

In this paper, we first limit ourself to focus our attention on the $%
\overline{K}N$ states $\Lambda (1405)$, $\Lambda (1600)$, $\Lambda (1670)$, $%
\Sigma (1385)$ and $\Sigma (1620)$ which were discussed in the previous
literature. Within the prescription of RGM, these states are naturally
treated as $\overline{K}N$ molecules with negative strangeness. The
calculated masses of these states as well as the experimental ones are
listed in Table 1. In our calculation, except for the color parameter $%
\alpha $, the other parameters are taken to be the same as used in the
scattering case. The color parameter which represents the color polarization
describes the color structure of a state which characterizes, to some
extent, the internal quark-gluon structure of the state. Obviously, the
parameter $\alpha $ in bound states would be different from that in
scattering processes. For the low energy elastic scattering, in general,
only peripheral interactions are concerned. Apparently, in this case, the
kaon (antikaon) and nucleon suffer from a slight color polarization in the
interacting region and in the initial state and final state, they are, as
free particles, still in color singlets. In our test, we find, when we take
the combination parameter $\alpha =0.915$, which means that the color
singlet interaction between kaon (antikaon) and nucleon dominates, the
scattering data may be fairly reproduced. However, in bound states and
resonances, the kaon (antikaon) and nucleon are bounded together and even
overlapped. In this case, a strong color polarization would happen. That is
to say, the kaon (antikaon) and nucleon in the bound states and resonances,
in general, can not exist as color singlet objects. Since the color
structure or the quark-gluon structures for different bound states and
resonances are different, the effective interactions taking place in those
states would be different from one another. According to our calculation,
the $\Lambda (1405)$ is identified with a lowest $\overline{K}N$ bound state
with the quantum numbers $S_{0\ 1}(\frac 12^{-})$ and color parameter $%
\alpha =0.91$ which implies that the both clusters $\overline{K}(\overline{q}%
s)$ and $N(q^3)$ in the $s$-shell state are almost kept in their color
singlets. While, the $\Lambda (1600)$, $\Lambda (1670)$, $\Sigma (1385)$ and 
$\Sigma (1620)$ are excited molecular states with quantum numbers $P_{0\ 1}(%
\frac 12^{+})^{*}$, $S_{0\ 1}(\frac 12^{-})^{*}$, $P_{1\ 3}(\frac 32^{+})^{*}
$ and $S_{1\ 1}(\frac 12^{-})^{*}$, respectively. The state $\Lambda (1600)$
is a $p$-shell resonance and the $\Lambda (1670)$ is a first radial-excited $%
s$-shell resonance both of which are isoscalar states. While, $\Sigma (1385)$
and $\Sigma (1620)$ are the isovectorial $p$-shell and $s$-shell states. The
smaller values of the parameter $\alpha $ for these states indicate that
they have rather complicated quark-gluon structures. Particularly, for the
resonance state $\Sigma (1385)$, since the $\alpha $ is too small, there is
a small amount of singlet-singlet component in this state and, therefore, it
cannot be viewed as an ordinary meson-baryon molecule. Such a state looks
like an exotic baryon state containing four valence quarks and one valence
antiquarks. As we learn from particle physics, colored quark and/or
antiquark pairs and gluons as well as their colored clusters existing in the
intermediate state of a process or in the interaction region of an hadron
system will decay into various color singlet hadrons in the final state.
Therefore, working in the hadron dynamics, to investigate the problems of
decay, scattering and bound states for hadron systems, it is natural to
introduce various meson exchange potentials and to perform coupled channel
calculations where all hadrons involved are treated from beginning to end as
color singlet objects. Nevertheless, when we work in the quark potential
model and introduce the color octet mechanism, consideration of the meson
exchanges may be unnecessary.

%TCIMACRO{\TeXButton{B}{\begin{table}[tbp] \centering} }
%BeginExpansion
\begin{table}[tbp] \centering
%EndExpansion
\begin{tabular}{llllll}
States & $\Lambda (1405)$ & $\Lambda (1600)$ & $\Lambda (1670)$ & $\Sigma
(1385)$ & $\Sigma (1620)$ \\ \hline
$L_{I\ 2J}(J^P)$ & $S_{0\ 1}(\frac 12^{-})$ & $P_{0\ 1}(\frac 12^{+})^{*}$ & 
$S_{0\ 1}(\frac 12^{-})^{*}$ & $P_{1\ 3}(\frac 32^{+})^{*}$ & $S_{1\ 1}(%
\frac 12^{-})^{*}$ \\ 
$\alpha $ & 0.91 & 0.83 & 0.723 & 0.21 & 0.81 \\ 
Theor. masses (MeV) & 1405 & 1603 & 1667 & 1384 & 1617 \\ 
Expt. masses (MeV) & 1405$\pm $10 & 1596$\pm $6 & 1670$\pm $5 & 1385$\pm $3
& 1633$\pm $10
\end{tabular}

\caption{Masses of exotic states S=-1\label{key}}%
%TCIMACRO{\TeXButton{E}{\end{table}}}
%BeginExpansion
\end{table}
%EndExpansion

Now let us turn to the resonance $\Theta ^{+}(1540)$. Whether this state
exists or not nowadays is still in debate. Some experiments supported its
existence, but some other experiments failed to find it [28]. In this paper,
we have calculated the $KN$ states by the same procedure as for the $%
\overline{K}N$ states, trying to find if a resonance could be formed in the $%
P_{01}$ state or the $D_{03}$ state. In the calculation, we use the
parameters as determined in the study of $KN$ scattering except that the
parameter $\alpha $ is chosen to be a adjustable parameter. But, in our
test, we find, even though in these two states, as mentioned before, the
interactions are attractive, it is impossible to find an appropriate $\alpha 
$ which could give a state with the mass $1540MeV\ $and $IJ^P=0\frac 12^{+}$
[37] or $IJ^P=0\frac 32^{-}$[38, 39]. The reason is probably due to that the
attractive interactions in those states are too weak for forming a bound
state or a resonance.

\section{Summary}

In this paper, the states $\Lambda (1405),$ $\Lambda (1600)$, $\Lambda
(1670) $, $\Sigma (1385)$, $\Sigma (1620)$ and $\Theta ^{+}(1540)$ were
investigated in the constituent quark potential model within the framework
of RGM. The distinctive feature of the investigation is that the effective
interaction potential between the strange meson and nucleon was merely
derived from QCD-inspired interquark potential with incorporating the QCD
renormalization correction and the spin-orbital suppression effect in it
without concerning any meson exchanges. With considering the contribution
arising from the color octets of the clusters $\overline{K}(\overline{q}s)$
(or $K(q\overline{s}))$ and $N(q^3),$ the potential model used is able to
reasonably reproduce the $KN$ low energy scattering data and gives some
prediction of the $\overline{K}N$ low energy elastic scattering phase
shifts. With the theoretical parameters are determined by fitting the
scattering data, the model calculation allows us to interpret the states $%
\Lambda (1405),$ $\Lambda (1600)$, $\Lambda (1670)$, $\Sigma (1385)$ and $%
\Sigma (1620)$ as $\overline{K}N$ molecular states with a certain color
structures characterized by the parameter $\alpha $ and gives no support to
the existence of the pentaquark state $\Theta ^{+}(1540)$. Certainly, the
validity of the model used in this paper needs to be verified by further
investigations on other hadron systems and on the problem of resonance
decays. In particular, to confirm the color structures of the $\overline{K}N$
molecular states, it is expected that more accurate lattice QCD calculations
of the states would appear in the future. Moreover, to justify the results
given in our calculation, experimental $\overline{K}N$ elastic scattering
phase shifts given in the low energy domain are urgently anticipated.

\section{Acknowledgment}

This project was supported in part by the National Natural Science
Foundation of China and Youth Foundation of Jilin University.

\section{Appendix A: The color-flavor-spin wave functions}

In general, the color singlet color state of the five quark cluster $(q^4%
\overline{s})$ or $(q^3\overline{q}s)$ may be built up by the color singlets
of the $N$-cluster $(qqq)$ and $K$-cluster $(q\overline{s})$ (or the $%
\overline{K}$-cluster $(\overline{q}s)$) or the color octets of the two
subclusters. Correspondingly, for the five quark cluster, there are two
classes of color-flavor-spin wave functions denoted by $\Psi _{TM\frac 12%
m}^{(1)}(1,2,3,4,5)$ and $\Psi _{TM\frac 12m}^{(2)}(1,2,3,4,5)$ which are
color singlets as a whole, but associated respectively with the color
singlets and the color octets of the two subclusters. In the function $\Psi
_{TM\frac 12m}^{(1)}(1,2,3,4,5),$ the color-flavor-spin (CFS) wave function $%
\Psi _{\frac 12M_1\frac 12m_s}^{(1)}(1,2,3)_N$ for the $N$-cluster which is
of the symmetry denoted by the Young diagram $[1^3]_{cfs}$ and hence totally
antisymmetric ) is constructed from the C-G coupling of $[1^3]_C\times
[3]_{FS}$ where $[1^3]_C$ and $[3]_{FS}$ are the Young diagrams denoting the
antisymmetric color singlet and the symmetric flavor-spin states
respectively. In the function $\Psi _{TM\frac 12m}^{(2)}(1,2,3,4,5)$, the
antisymmetric CFS wave function $\Psi _{\frac 12M_1\frac 12%
m_s}^{(1)}(1,2,3)_N$ for the $N$-cluster is given by the C-G coupling of $%
[21]_C\times [21]_{FS}$ where $[21]_C$ and $[21]_{FS}$ represent the color
octet state and the flavor-spin state of mixed symmetry respectively. The
explicit expressions of the wave functions mentioned above can easily be
written out by the familiar method given in the group theory, as displayed
in the following.

The first class of the CFS wave function in Eq. (12) for the whole system is

\begin{equation}
\Psi _{TM\frac 12m}^{(1)}(1,2,3,4,5)=\sum\limits_{M_1M_2}C_{\frac 12M_1\frac 
12M_2}^{TM}\Psi _{\frac 12M_1\frac 12m}^{(1)}(1,2,3)_N\Psi _{\frac 12%
M_200}^{(1)}(4,5)_K  \eqnum{A1}
\end{equation}
where $\Psi _{\frac 12M_1\frac 12m}^{(1)}(1,2,3)_N$ , as mentioned before,
is the CFS wave function for the $N$-cluster and $\Psi _{\frac 12%
M_200}^{(1)}(4,5)_K$ is the CFS wave function for the $K$-cluster. They are
represented separately as 
\begin{equation}
\Psi _{\frac 12M_1\frac 12m_s}^{(1)}(1,2,3)_N=\xi _c^0(1,2,3)\chi _{\frac 12%
M_1\frac 12m_s}^{(1)}(1,2,3)  \eqnum{A2}
\end{equation}
where 
\begin{equation}
\xi _c^0(1,2,3)=\frac 1{\sqrt{6}}\epsilon _{abc}q^a(1)q^b(2)q^c(3) 
\eqnum{A3}
\end{equation}
represents the color singlet wave function of the $N$-cluster and 
\begin{equation}
\chi _{\frac 12M_1\frac 12m_s}^{(1)}(1,2,3)=\frac 1{\sqrt{2}}[\chi _{\frac 12%
M_1}^a(1,2,3)\varphi _{\frac 12m_s}^a(1,2,3)+\chi _{\frac 12%
M_1}^b(1,2,3)\varphi _{\frac 12m_s}^b(1,2,3)]  \eqnum{A4}
\end{equation}
is the isospin-spin wave function of the $N$-cluster in which the isospin
wave functions $\chi _{\frac 12M_1}^a(1,2,3)$ and $\chi _{\frac 12%
M_1}^b(1,2,3)$ and the spin wave functions $\varphi _{\frac 12m_s}^a(1,2,3)$
and $\varphi _{\frac 12m_s}^a(1,2,3)$ are expressed as follows 
\begin{equation}
\begin{tabular}{l}
$\chi _{\frac 12M_1}^a(1,2,3)=\sum\limits_{m,m_{3,}m_1,m_2}C_{1m\frac 12%
m_3}^{\frac 12M_1}C_{\frac 12m_1\frac 12m_2}^{1m}\chi _{\frac 12m_1}(1)\chi
_{\frac 12m_2}(2)\chi _{\frac 12m_3}(3)$ \\ 
$\chi _{\frac 12M_1}^b(1,2,3)=\sum\limits_{m_1,m_2,m_3}C_{00\frac 12M_1}^{%
\frac 12M_1}C_{\frac 12m_1\frac 12m_2}^{00}\chi _{\frac 12m_1}(1)\chi _{%
\frac 12m_2}(2)\chi _{\frac 12m_3}(3)$ \\ 
$\varphi _{\frac 12m_s}^a(1,2,3)=\sum\limits_{m,m_3,m_1,m_2}C_{1m\frac 12%
m_3}^{\frac 12m_s}C_{\frac 12m_1\frac 12m_2}^{1m}\varphi _{\frac 12%
m_2}(1)\varphi _{\frac 12m_2}(2)\varphi _{\frac 12m_3}(3)$ \\ 
$\varphi _{\frac 12m_s}^b(1,2,3)=\sum\limits_{m_1,m_2,m_3}C_{00\frac 12m_s}^{%
\frac 12m_s}C_{\frac 12m_1\frac 12m_2}^{00}\varphi _{\frac 12m_2}(1)\varphi
_{\frac 12m_2}(2)\varphi _{\frac 12m_3}(3)$%
\end{tabular}
\eqnum{A5}
\end{equation}
The CFS wave function of the $K$- cluster is 
\begin{equation}
\Psi _{\frac 12M00}^{(1)}(4,5)_K=C_0(4,5)\chi _{\frac 12M}(4,5)\varphi
_{00}(4,5)  \eqnum{A6}
\end{equation}
where $C_0(4,5)$, $\chi _{\frac 12M}(4,5)$ and $\varphi _{00}(4,5)$ are the
color, isospin and spin wave functions, respectively. Since there is no
identical particles in the cluster, these wave functions are of the forms 
\begin{equation}
C_0(4,5)=\frac 1{\sqrt{3}}q^a(4)\overline{q}^a(5)  \eqnum{A7}
\end{equation}
and 
\begin{equation}
\begin{tabular}{l}
$\chi _{\frac 12M}(4,5)=\sum\limits_{m_1\text{ }m_2}C_{\frac 12m_100}^{\frac 
12M}\chi _{\frac 12m_1}(4)\chi _{00}(5)$ \\ 
$\varphi _{00}(4,5)=\sum\limits_{m_1\text{ }m_2}C_{\frac 12m_1\frac 12%
m_2}^{00}\varphi _{\frac 12m_1}(4)\varphi _{\frac 12m_2}(5)$%
\end{tabular}
\eqnum{A8}
\end{equation}
For the second class of the CFS wave function in Eq. (12), it can be
represented as 
\begin{equation}
\Psi _{TM\frac 12m}^{(2)}(1,2,3,4,5)=\sum\limits_{M_1M_2}\sum\limits_cC_{%
\frac 12M_1\frac 12M_2}^{TM_T}\Psi _{\frac 12M_1\frac 12m}^{(2)c}(1,2,3)_N%
\Psi _{\frac 12M00}^{(2)c}(4,5)_K  \eqnum{A9}
\end{equation}
where $\Psi _{\frac 12M_1\frac 12m}^{(2)c}(1,2,3)_N$ and $\Psi _{\frac 12%
M00}^{(2)c}(4,5)_K$ are the second class of CFS wave functions for the $N$%
-cluster and the $K$-cluster respectively. Their expressions are shown in
the following. 
\begin{equation}
\Psi _{\frac 12M_1\frac 12m_s}^{(2)C}(1,2,3)_N=\frac 1{\sqrt{2}}[\xi
_c^A(1,2,3)\chi _{\frac 12M_1\frac 12m_s}^{(2)B}(1,2,3)-\xi _c^B(1,2,3)\chi
_{\frac 12M_1\frac 12m_s}^{(2)A}(1,2,3)]  \eqnum{A10}
\end{equation}
where $\xi _c^A(1,2,3)$ and $\xi _c^B(1,2,3)$ are the color octet wave
functions given respectively by the Young-Tableau [211] and the
Young-Tableau [121] and $\chi _{\frac 12M_1\frac 12m_s}^{(2)A}(1,2,3)$ and $%
\chi _{\frac 12M_1\frac 12m_s}^{(2)B}(1,2,3)$ are the corresponding
isospin-spin wave functions. Their expressions are 
\begin{equation}
\begin{tabular}{l}
$\xi _c^A(1,2,3)=\frac 12\epsilon
_{ijb}[q^a(1)q^i(2)q^j(3)+q^a(2)q^i(1)q^j(3)]$ \\ 
$\xi _c^B(1,2,3)=\frac 1{2\sqrt{3}}\epsilon
_{ijb}[q^a(1)q^i(2)q^j(3)-q^a(2)q^i(1)q^j(3)-2\text{ }q^a(3)q^i(1)q^j(2)]$
\\ 
$\chi _{\frac 12M_1\frac 12m_s}^{(2)A}(1,2,3)=\frac 1{\sqrt{2}}[\chi _{\frac 
12M_1}^a(1,2,3)\varphi _{\frac 12m_s}^a(1,2,3)-\chi _{\frac 12%
M_1}^b(1,2,3)\varphi _{\frac 12m_s}^b(1,2,3)]$ \\ 
$\chi _{\frac 12M_1\frac 12m_s}^{(2)B}(1,2,3)=-\frac 1{\sqrt{2}}[\chi _{%
\frac 12M_1}^a(1,2,3)\varphi _{\frac 12m_s}^b(1,2,3)+\chi _{\frac 12%
M_1}^b(1,2,3)\varphi _{\frac 12m_s}^a(1,2,3)]$%
\end{tabular}
\eqnum{A11}
\end{equation}
The second class of the CFS wave function for the $K$(or $\overline{K}$%
)-cluster is as follows 
\begin{equation}
\Psi _{\frac 12M00}^{(2)c}(4,5)_K=C_a^b(4,5)\chi _{\frac 12M}(4,5)\varphi
_{00}(4,5)  \eqnum{A12}
\end{equation}
where 
\begin{equation}
C_a^b(4,5)=q^b(4)q_a(5)-\frac 13\delta _a^bq^c(4)q_c(5)  \eqnum{A13}
\end{equation}
is the color octet for the $K(\overline{K})$ cluster and the other two
functions $\chi _{\frac 12M}(4,5)$, $\varphi _{00}(4,5)$ are the same as in
(A8).

\section{Appendix B: The effective $KN$ and $\overline{K}N$ interaction
potentials}

In this appendix, we show the nonlocal effective interaction potentials of
the $KN$ and $\overline{K}N$ systems which are derived from the interquark
potentials written in section 2 by the resonating group approach.

The $KN$ nonlocal effective potential $V_t(\vec R,\vec R^{\prime })$ which
is derived from the $t$-channel OGEP written in Eq. (4) is divided into two
parts: the direct part $V_t^D(\vec R,\vec R^{\prime })$ and the exchanged
part $V_t^{ex}(\vec R,\vec R^{\prime })$:

\begin{equation}
V_t(\vec R,\vec R^{\prime })=V_t^D(\vec R,\vec R^{\prime })-V_t^{ex}(\vec R,%
\vec R^{\prime })  \eqnum{B1}
\end{equation}
where 
\begin{equation}
V_t^{ex}(\vec R,\vec R^{\prime })=V_t^{ex}(\vec R,\vec R^{\prime
})^{14}+V_t^{ex}(\vec R,\vec R^{\prime })^{24}+V_t^{ex}(\vec R,\vec R%
^{\prime })^{34}  \eqnum{B2}
\end{equation}
here the superscript $ab=14,24$ or $34$ designates which pair of quarks
interchange. Each part of the potential contains several terms as follows: 
\begin{equation}
\begin{tabular}{l}
$V_t^D(\vec R,\vec R^{\prime })=V_{15}^D(\vec R,\vec R^{\prime })+V_{25}^D(%
\vec R,\vec R^{\prime })+V_{35}^D(\vec R,\vec R^{\prime })$ \\ 
$+V_{14}^D(\vec R,\vec R^{\prime })+V_{24}^D(\vec R,\vec R^{\prime
})+V_{34}^D(\vec R,\vec R^{\prime })$%
\end{tabular}
\eqnum{B3}
\end{equation}
and 
\begin{equation}
\begin{tabular}{l}
$V_t^{ex}(\vec R,\vec R^{\prime })^{ab}=V_{14}^{ex}(\vec R,\vec R^{\prime
})^{ab}+V_{24}^{ex}(\vec R,\vec R^{\prime })^{ab}+V_{34}^{ex}(\vec R,\vec R%
^{\prime })^{ab}+V_{12}^{ex}(\vec R,\vec R^{\prime })^{ab}+V_{23}^{ex}(\vec R%
,\vec R^{\prime })^{ab}$ \\ 
$+V_{15}^{ex}(\vec R,\vec R^{\prime })^{ab}+V_{25}^{ex}(\vec R,\vec R%
^{\prime })^{ab}+V_{35}^{ex}(\vec R,\vec R^{\prime })^{ab}+V_{45}^{ex}(\vec R%
,\vec R^{\prime })^{ab}+V_{13}^{ex}(\vec R,\vec R^{\prime })^{ab}$%
\end{tabular}
\eqnum{B4}
\end{equation}
where the subscript in each term on the right hand sides (RHS) of (B3) and
(B4) marks the two interacting quarks: one in the $N$-cluster, another in
the $K$-cluster. According to Eq. (25), the terms $V_{ij}^D(\vec R,\vec R%
^{\prime })$ and $V_{ij}^{ex}(\vec R,\vec R^{\prime })^{ab}$ are derived in
such a way 
\begin{equation}
\begin{array}{c}
V_{ij}^D(\vec R,\vec R^{\prime })=\int d^3\rho d^3\rho ^{^{\prime }}\Gamma (%
\vec R,\vec \rho )V_{ij}^D(\vec \rho ,\vec \rho ^{\prime })\Gamma (\vec \rho
^{\prime },\vec R^{\prime }), \\ 
V_{ij}^{ex}(\vec R,\vec R^{\prime })^{ab}=\int d^3\rho d^3\rho ^{^{\prime
}}\Gamma (\vec R,\vec \rho )V_{ij}^{ex}(\vec \rho ,\vec \rho ^{\prime
})^{ab}\Gamma (\vec \rho ^{\prime },\vec R^{\prime })
\end{array}
\eqnum{B5}
\end{equation}
in which $V_{ij}^D(\vec \rho ,\vec \rho ^{\prime })$ and $V_{ij}^{ex}(\vec 
\rho ,\vec \rho ^{\prime })^{ab}$ can be explicitly calculated according to
the definition given in Eq. (20). It should be emphasized that the effective
potential in Eq. (23) is defined by the matrix element between the basis
wave function written in Eq. (10). In order to exhibit the spin, color and
isospin structure of the effective potential, we would like here to give the
effective potential in the operator form. The potential operator is derived
from the matrix element of the quark potential in Eq. (3) between the
position space wave function only, as illustrated in the following:

\begin{equation}
\hat V_{ij}^D(\vec \rho ,\vec \rho ^{\prime })=\int \prod_{i=1}^5\frac{d^3p_k%
}{(2\pi )^3}\frac{d^3p_k^{\prime }}{(2\pi )^3}\langle R(\vec p_1,\vec p_2,%
\vec p_3,\vec p_4,\vec p_5;\vec \rho )\mid V_{ij}^t\mid R(\vec p_1,\vec p_2,%
\vec p_3,\vec p_4,\vec p_5;\vec \rho ^{\prime })\rangle  \eqnum{B6}
\end{equation}
and 
\begin{equation}
\hat V_{ij}^{ex}(\vec \rho ,\vec \rho ^{\prime })^{ab}=\int \prod_{i=1}^5%
\frac{d^3p_k}{(2\pi )^3}\frac{d^3p_k^{\prime }}{(2\pi )^3}\langle R(\vec p_1,%
\vec p_2,\vec p_3,\vec p_4,\vec p_5;\vec \rho )\mid V_{ij}^tP_{ab}\mid R(%
\vec p_1,\vec p_2,\vec p_3,\vec p_4,\vec p_5;\vec \rho ^{\prime })\rangle 
\eqnum{B7}
\end{equation}
where $V_{ij}^t$ is $t$-channel OGEP represented in Eq. (4) and $R(\vec p_1,%
\vec p_2,\vec p_3,\vec p_4,\vec p_5;\vec \rho )$ is the position space wave
function written in Eq. (14). Clearly, the expressions written in Eq. (24)
and ((B1)-(B5) are formally kept unchanged when the potentials in those
expressions are replaced by the corresponding operator ones.{\it \ }For
instance, when the functions $V_{ij}^D(\vec \rho ,\vec \rho ^{\prime })$ and 
$V_{ij}^{ex}(\vec \rho ,\vec \rho ^{\prime })^{ab}$ in (B5) are replaced by
the corresponding operators $\hat V_{ij}^D(\vec \rho ,\vec \rho ^{\prime })$
and $\hat V_{ij}^{ex}(\vec \rho ,\vec \rho ^{\prime })^{ab}$ defined in (B6)
and (B7), through tedious calculations, one may obtain the potential
operators $\hat V_{ij}^D(\vec R,\vec R^{\prime })$ and $\hat V_{ij}^{ex}(%
\vec R,\vec R^{\prime })^{ab}$. The calculated results are displayed below.

First we describe the potential operators which correspond to the ten terms
on the RHS of (B4). By introducing the following functions: 
\begin{equation}
\begin{array}{c}
f_1^t(\vec R,\vec R^{\prime })_{ex}=\exp \{-\frac \zeta {4b^2}[(2\zeta -1)(%
\vec R-\vec R^{\prime })^2+\frac 1{2\zeta -1}(\vec R+\vec R^{\prime })^2]\}
\\ 
f_2^t(\vec R,\vec R^{\prime })_{ex}=\exp \{-\frac \gamma {2(3\zeta -2)b^2}%
[(4\zeta ^2-3\zeta +2)\vec R^2-8\zeta (\zeta -1)\vec R\cdot \vec R^{\prime
}+(4\zeta ^2-5\zeta +2)\vec R^{\prime 2}]\} \\ 
f_3^t(\vec R,\vec R^{\prime })_{ex}=\exp \{-\frac{\zeta (2\zeta -1)}{4b^2}(%
\vec R-\vec R^{\prime })^2-\frac{\zeta \cdot (1/2+\alpha _2)}{4b^2(\zeta
-1/2-\alpha _2)}(\vec R+\vec R^{\prime })^{^2}\} \\ 
f_4^t(\vec R,\vec R^{\prime })_{ex}=\exp \{-\frac \zeta {2(2\ \zeta -1-\zeta
\ \alpha _2)b^2}\{[\zeta (\zeta +\alpha _2)+(\zeta -1)^2]\vec R^2+4\zeta
(\zeta -1)\vec R\cdot \vec R^{\prime } \\ 
+[\zeta (\zeta -\alpha _2)+(\zeta -1)^2]\ \vec R^{\prime 2}\}\}
\end{array}
\eqnum{B8}
\end{equation}
with $\zeta =3\beta _2$, the exchanged terms of the potential operator $\hat 
V_t^{ex}(\vec R,\vec R^{\prime })^{14}$ can be written as 
\begin{equation}
\begin{array}{c}
\hat V_{24}^{ex}(\vec R,\vec R^{\prime })^{14}=\frac{16\alpha
_sC_{24}^t\zeta ^3}{\pi ^2b^4}(\frac \zeta {2(2\zeta -1)})^{\frac 32}f_1^t(%
\vec R,\vec R^{\prime })_{ex}\{1-\frac 1{4m_1^2b^2}[\frac 94+3(2\beta
_2-1/2)^2+\frac{\frac 32\zeta }{2\zeta -1}+ \\ 
\frac{\zeta ^2}{2b^2}\frac{4\beta _2-1}{2\zeta -1}(\vec R^2-\vec R^{\prime
2})-\frac{\zeta ^2}{4b^2(2\zeta -1)^2}(\vec R+\vec R^{\prime })^2-\frac{%
\zeta ^2}{b^2}(2\beta _2-\frac 12)^2(\vec R-\vec R^{\prime })^2] \\ 
-\frac{\zeta ^2}{4m_1^2b^4(2\zeta -1)^2}[(\zeta -1)\vec R^{\prime }-\zeta 
\vec R]^2+\frac{3\zeta }{8m_1^2b^2(2\zeta -1)}+\frac 3{4m_1^2b^2}-i\frac{%
\zeta ^2}{4m_1^2b^4(2\zeta -1)}(1+\gamma )(\vec R\times \vec R^{\prime
})\cdot (\vec \sigma _2-\vec \sigma _4) \\ 
+\frac 1{4m_1^2}[(\frac \gamma {b^2}+\frac{\gamma ^2}{4b^2}\frac{2\zeta }{%
2\zeta -1})\vec \sigma _2\cdot \vec \sigma _4-\frac{\gamma ^2\zeta ^2}{%
4b^4(2\zeta -1)^2}(\zeta \vec R-(\zeta -1)\vec R^{\prime })\cdot \vec \sigma
_2(\zeta \vec R-(\zeta -1)\vec R^{\prime })]\} \\ 
-\frac{4\alpha _sC_{24}^t\gamma ^3}{\pi ^2m_1^2b^6}(\frac \zeta {3\zeta -1}%
)^{\frac 32}f_2^t(\vec R,\vec R^{\prime })_{ex}(1+\vec \sigma _2\cdot \vec 
\sigma _4),
\end{array}
\eqnum{B9}
\end{equation}
\begin{equation}
\begin{array}{c}
\hat V_{25}^{ex}(\vec R,\vec R^{\prime })^{14}=\frac{8\alpha _sC_{25}^t\zeta
^3}{\alpha _2\pi ^2b^4}(\frac{\beta _1}{2\zeta -1})^{\frac 32}f_1^t(\vec R,%
\vec R^{\prime })_{ex}\{1-\frac 1{m_{12}^2b^2}[\frac 3{2\alpha _1}+3\frac{%
(\alpha _1-\beta _1)^2}{\alpha _1^2}-\frac{(\alpha _1-\beta _1)^2\zeta ^2}{%
\alpha _1^2b^2}(\vec R-\vec R^{\prime })^2] \\ 
+\frac{m_1-m_2}{m_1m_2m_{12}b^2}[\frac{3(\alpha _1-\beta _1)}\alpha -\frac{%
(\alpha _1-\beta _1)\ \zeta ^2}{\alpha \ b^2}(\vec R-\vec R^{\prime })^2]-i%
\frac{(\alpha _1-\beta _1)\ \zeta ^2}{2m_{12}\alpha \ b^4(2\zeta -1)}%
(1+\gamma )(\vec R\times \vec R^{\prime })\cdot (\frac{\vec \sigma _2}{m_1}-%
\frac{\vec \sigma _5}{m_2}) \\ 
+i\frac{(\alpha _1-\beta _1)\ \zeta ^2\alpha _2^2}{2m_1m_2\ b^4(2\zeta -1)}%
(1+\gamma )(\vec R\times \vec R^{\prime })\cdot [(2+\frac{m_2}{m_1})\vec 
\sigma _2-(2+\frac{m_1}{m_2})\vec \sigma _5] \\ 
+\frac 1{4m_1m_2}[(\frac{2\alpha _2\gamma }{b^2}+\frac{\gamma ^2}{b^2}\frac{%
4\alpha _2^2}{2\zeta -1})\vec \sigma _2\cdot \vec \sigma _5-\frac{4\gamma
^2\zeta ^2\alpha _2^2}{b^4(2\zeta -1)^2}(\vec R+\vec R^{\prime })\cdot \vec 
\sigma _2(\vec R+\vec R^{\prime })\cdot \vec \sigma _5]\} \\ 
-\frac{16\alpha _sC_{25}^t\zeta ^3}{\pi ^2b^6}(\frac{\beta _1}{2(\zeta -%
\frac 12-\alpha _2)})^{\frac 32}f_3^t(\vec R,\vec R^{\prime })_{ex}\frac 1{%
4m_1m_2}(\frac{m_1^2+m_2^2}{2m_1m_2}+\vec \sigma _2\cdot \vec \sigma _5),
\end{array}
\eqnum{B10}
\end{equation}
\begin{equation}
\begin{array}{c}
\hat V_{15}^{ex}(\vec R,\vec R^{\prime })^{14}=\frac{8\alpha _sC_{15}^t\zeta
^3}{\alpha _2\pi ^2b^4}(\frac{\beta _1}{2\zeta -1})^{\frac 32}f_1^t(\vec R,%
\vec R^{\prime })_{ex}\{1-\frac 1{m_{12}^2b^2}[\frac \zeta {2\alpha _1}+%
\frac{3\ \zeta \ }2(2\ \zeta -1+\frac 1{2\ \zeta -1}) \\ 
-\frac{\zeta ^2}{4(2\ \zeta -1)^2b^2}([(2\ \zeta -1)^2-1]\vec R-[(2\ \zeta
-1)^2+1]\vec R^{\prime })^2]+\frac{\alpha _2(m_1-m_2)}{4m_1m_2m_{12}b^2(2\
\zeta -1)}[-12\zeta (\zeta -1) \\ 
-\frac{2\ \zeta ^2}{b^2(2\ \zeta -1)}((\zeta -1)\vec R-\zeta \vec R^{\prime
})([(2\zeta -1)^2-1]\vec R-[(2\zeta -1)^2+1]\vec R^{\prime })]+\frac 1{%
4m_1m_2b^2}[6\alpha _2(1+\frac{\alpha _2\zeta }{2\zeta -1}) \\ 
-\frac{4\zeta ^2\alpha _2^2}{b^2(2\ \zeta -1)^2}((2\zeta +1)\vec R-(2\zeta
-3)\vec R^{\prime })^2]+i\frac{\beta _1\zeta ^2}{2m_{12}\alpha \ b^4(2\zeta
-1)}(1+\gamma )(\vec R\times \vec R^{\prime })\cdot (\frac{\vec \sigma _1}{%
m_1}-\frac{\vec \sigma _5}{m_2}) \\ 
+\frac 1{4m_1m_2}[(\frac{2\alpha _2\gamma }{b^2}+\frac{\gamma ^2}{b^2}\frac{%
2\zeta \alpha _2^2}{2\zeta -1})\vec \sigma _1\cdot \vec \sigma _5-\frac{%
4\gamma ^2\zeta ^2\alpha _2^2}{b^4(2\zeta -1)^2}(\varsigma \vec R-(\zeta -1)%
\vec R^{\prime })\cdot \vec \sigma _1(\varsigma \vec R-(\zeta -1)\vec R%
^{\prime })\cdot \vec \sigma _5]\} \\ 
-\frac{16\alpha _sC_{15}^t\zeta ^3}{\pi ^2b^6}(\frac{\beta _1}{2\zeta
-1-\zeta \alpha _2)})^{\frac 32}f_4^t(\vec R,\vec R^{\prime })_{ex}\frac 1{%
4m_1m_2}(\frac{m_1^2+m_2^2}{2m_1m_2}+\vec \sigma _1\cdot \vec \sigma _5),
\end{array}
\eqnum{B11}
\end{equation}
\begin{equation}
\begin{array}{c}
\hat V_{14}^{ex}(\vec R,\vec R^{\prime })^{14}=\frac{16\alpha
_sC_{14}^t\zeta ^3}{\pi ^2b^4}(\frac \zeta {2(2\zeta -1)})^{\frac 32}f_1^t(%
\vec R,\vec R^{\prime })_{ex}\{1-\frac 1{4m_1^2b^2}[3+3(\beta _2-\beta _1)^2-%
\frac{\zeta ^2(\beta _2-\beta _1)^2}{b^2}(\vec R-\vec R^{\prime })^2] \\ 
+\frac 1{4m_1^2b^2}[3+\frac{12}{2\zeta -1}-\frac{4\ \zeta ^2}{b^2(2\zeta
-1)^2}(\vec R+\vec R^{\prime })^2]-i\frac{3\zeta ^2}{4m_1^2b^4(2\zeta -1)}%
(1+\gamma )(\vec R\times \vec R^{\prime })\cdot (\vec \sigma _1+\vec \sigma
_4) \\ 
+\frac 1{4m_1^2b^2}[(\gamma +\gamma ^2)\vec \sigma _1\cdot \vec \sigma _4-%
\frac{\gamma ^2\zeta ^2}{b^2}(\vec R-\vec R^{\prime })\cdot \vec \sigma _1(%
\vec R-\vec R^{\prime })\cdot \vec \sigma _4]\} \\ 
-\frac{\alpha _sC_{14}^t}{\pi ^2m_1^2b^3}(\frac{\pi \ \zeta }{2\zeta -1})^{%
\frac 32}\delta (\vec R-\vec R^{\prime })e^{-\frac \zeta {4\ b^2(2\zeta -1)}(%
\vec R+\vec R^{\prime })^2}(1+\vec \sigma _1\cdot \vec \sigma _4),
\end{array}
\eqnum{B12}
\end{equation}
\begin{equation}
\begin{array}{c}
\hat V_{34}^{ex}(\vec R,\vec R^{\prime })^{14}=\frac{16\alpha
_sC_{34}^t\zeta ^3}{\pi ^2b^4}(\frac \zeta {2(2\zeta -1)})^{\frac 32}f_1^t(%
\vec R,\vec R^{\prime })_{ex}\{1-\frac 1{4m_1^2b^2}[3+3(\beta _2-\beta _1)^2-%
\frac{\zeta ^2(\beta _2-\beta _1)^2}{b^2}(\vec R-\vec R^{\prime })^2] \\ 
+\frac 1{4m_1^2b^2}[3+\frac{12}{2\zeta -1}-\frac{4\ \zeta ^2}{b^2(2\zeta
-1)^2}(\vec R+\vec R^{\prime })^2]-i\frac{3\zeta ^2}{4m_1^2b^4(2\zeta -1)}%
(1+\gamma )(\vec R\times \vec R^{\prime })\cdot (\vec \sigma _3+\vec \sigma
_4) \\ 
+\frac 1{4m_1^2b^2}[(\gamma +\gamma ^2)\vec \sigma _3\cdot \vec \sigma _4-%
\frac{\gamma ^2\zeta ^2}{b^2}(\vec R-\vec R^{\prime })\cdot \vec \sigma _3(%
\vec R-\vec R^{\prime })\cdot \vec \sigma _4]\} \\ 
-\frac{\alpha _sC_{34}^t}{\pi ^2m_1^2b^3}(\frac{\pi \ \zeta }{2\zeta -1})^{%
\frac 32}\delta (\vec R-\vec R^{\prime })e^{-\frac \zeta {4\ b^2(2\zeta -1)}(%
\stackrel{\rightharpoonup }{R}+\stackrel{\rightharpoonup }{R^{\prime }}%
)^2}(1+\vec \sigma _3\cdot \vec \sigma _4)
\end{array}
\eqnum{B13}
\end{equation}
\begin{equation}
\begin{array}{c}
\hat V_{12}^{ex}(\vec R,\vec R^{\prime })^{14}=\frac{16\alpha
_sC_{12}^t\zeta ^3}{\pi ^2b^4}(\frac \zeta {2(2\zeta -1)})^{\frac 32}f_1^t(%
\vec R,\vec R^{\prime })_{ex}\{1-\frac 1{4m_1^2b^2}[3+3(\beta _2-\beta _1)^2-%
\frac{\zeta ^2(\beta _2-\beta _1)^2}{b^2}(\vec R-\vec R^{\prime })^2] \\ 
+\frac 1{4m_1^2b^2}[3+\frac{12}{2\zeta -1}-\frac{4\ \zeta ^2}{b^2(2\zeta
-1)^2}(\vec R+\vec R^{\prime })^2]+i\frac{3\zeta ^2}{4m_1^2b^4(2\zeta -1)}%
(1+\gamma )(\vec R\times \vec R^{\prime })\cdot (\vec \sigma _1+\vec \sigma
_2) \\ 
+\frac 1{4m_1^2b^2}[(\gamma +\gamma ^2)\vec \sigma _1\cdot \vec \sigma _2-%
\frac{\gamma ^2\zeta ^2}{b^2}(\vec R-\vec R^{\prime })\cdot \vec \sigma _1(%
\vec R-\vec R^{\prime })\cdot \vec \sigma _2]\} \\ 
-\frac{\alpha _sC_{12}^t}{\pi ^2m_1^2b^3}(\frac{\pi \ \zeta }{2\zeta -1})^{%
\frac 32}\delta (\vec R-\vec R^{\prime })e^{-\frac \zeta {4\ b^2(2\zeta -1)}(%
\vec R+\vec R^{\prime })^2}(1+\vec \sigma _1\cdot \vec \sigma _2),
\end{array}
\eqnum{B14}
\end{equation}
\begin{equation}
\begin{array}{c}
\hat V_{13}^{ex}(\vec R,\vec R^{\prime })^{14}=\frac{16\alpha
_sC_{13}^t\zeta ^3}{\pi ^2b^4}(\frac \zeta {2(2\zeta -1)})^{\frac 32}f_1^t(%
\vec R,\vec R^{\prime })_{ex}\{1-\frac 1{4m_1^2b^2}[3+3(\beta _2-\beta _1)^2-%
\frac{\zeta ^2(\beta _2-\beta _1)^2}{b^2}(\vec R-\vec R^{\prime })^2] \\ 
+\frac 1{4m_1^2b^2}[3+\frac{12}{2\zeta -1}-\frac{4\ \zeta ^2}{b^2(2\zeta
-1)^2}(\vec R+\vec R^{\prime })^2]+i\frac{3\zeta ^2}{4m_1^2b^4(2\zeta -1)}%
(1+\gamma )(\vec R\times \vec R^{\prime })\cdot (\vec \sigma _1+\vec \sigma
_3) \\ 
+\frac 1{4m_1^2b^2}[(\gamma +\gamma ^2)\vec \sigma _1\cdot \vec \sigma _3-%
\frac{\gamma ^2\zeta ^2}{b^2}(\vec R-\vec R^{\prime })\cdot \vec \sigma _1(%
\vec R-\vec R^{\prime })\cdot \vec \sigma _3]\} \\ 
-\frac{\alpha _sC_{13}^t}{\pi ^2m_1^2b^3}(\frac{\pi \ \zeta }{2\zeta -1})^{%
\frac 32}\delta (\vec R-\vec R^{\prime })e^{-\frac \zeta {4\ b^2(2\zeta -1)}(%
\vec R+\vec R^{\prime })^2}(1+\vec \sigma _1\cdot \vec \sigma _3),
\end{array}
\eqnum{B15}
\end{equation}
\begin{equation}
\begin{array}{c}
\hat V_{23}^{ex}(\vec R,\vec R^{\prime })^{14}=\frac{16\alpha
_sC_{23}^t\zeta ^3}{\pi ^2b^4}(\frac \zeta {2(2\zeta -1)})^{\frac 32}f_1^t(%
\vec R,\vec R^{\prime })_{ex}\{1-\frac 1{4m_1^2b^2}[3+12\ \beta _2^2 \\ 
-\frac{4\ \beta _2^2\zeta ^2}{b^2}(\vec R^2-\vec R^{\prime 2})]+\frac 1{%
2m_1^2b^2}-\frac 1{6m_1^2b^2}\vec \sigma _2\cdot \vec \sigma _3\},
\end{array}
\eqnum{B16}
\end{equation}

\begin{equation}
\begin{array}{c}
\hat V_{35}^{ex}(\vec R,\vec R^{\prime })^{14}=\frac{8\alpha _sC_{35}^t\zeta
^3}{\alpha _2\pi ^2b^4}(\frac{\beta _1}{2\zeta -1})^{\frac 32}f_1^t(\vec R,%
\vec R^{\prime })_{ex}\{1-\frac 1{m_{12}^2b^2}[\frac 3{2\alpha _1}+3\frac{%
(\alpha _1-\beta _1)^2}{\alpha _1^2}-\frac{(\alpha _1-\beta _1)^2\zeta ^2}{%
\alpha _1^2b^2}(\vec R-\vec R^{\prime })^2] \\ 
+\frac{m_1-m_2}{m_1m_2m_{12}b^2}[\frac{3(\alpha _1-\beta _1)}\alpha -\frac{%
(\alpha _1-\beta _1)\ \zeta ^2}{\alpha \ b^2}(\vec R-\vec R^{\prime })^2]-i%
\frac{(\alpha _1-\beta _1)\ \zeta ^2}{2m_{12}\alpha \ b^4(2\zeta -1)}%
(1+\gamma )(\vec R\times \vec R^{\prime })\cdot (\frac{\vec \sigma _3}{m_1}-%
\frac{\vec \sigma _5}{m_2}) \\ 
+i\frac{(\alpha _1-\beta _1)\ \zeta ^2\alpha _2^2}{2m_1m_2\ b^4(2\zeta -1)}%
(1+\gamma )(\vec R\times \vec R^{\prime })\cdot [(2+\frac{m_2}{m_1})\vec 
\sigma _3-(2+\frac{m_1}{m_2})\vec \sigma _5] \\ 
+\frac 1{4m_1m_2}[(\frac{2\alpha _2\gamma }{b^2}+\frac{\gamma ^2}{b^2}\frac{%
4\alpha _2^2}{2\zeta -1})\vec \sigma _3\cdot \vec \sigma _5-\frac{4\gamma
^2\zeta ^2\alpha _2^2}{b^4(2\zeta -1)^2}(\vec R+\vec R^{\prime })\cdot \vec 
\sigma _3(\vec R+\vec R^{\prime })\cdot \vec \sigma _5]\} \\ 
-\frac{16\alpha _sC_{35}^t\zeta ^3}{\pi ^2b^6}(\frac{\beta _1}{2(\zeta -%
\frac 12-\alpha _2)})^{\frac 32}f_3^t(\vec R,\vec R^{\prime })_{ex}\frac 1{%
4m_1m_2}(\frac{m_1^2+m_2^2}{2m_1m_2}+\vec \sigma _3\cdot \vec \sigma _5)
\end{array}
\eqnum{B17}
\end{equation}
and 
\begin{equation}
\begin{array}{c}
\hat V_{45}^{ex}(\vec R,\vec R^{\prime })^{14}=\frac{8\alpha _sC_{45}^t\zeta
^3}{\alpha _2\pi ^2b^4}(\frac{\beta _1}{2\zeta -1})^{\frac 32}f_1^t(\vec R,%
\vec R^{\prime })_{ex}\{1-\frac 1{m_{12}^2b^2}[\frac \zeta {2\alpha _1}+%
\frac{3\ \zeta \ }2(2\ \zeta -1+\frac 1{2\ \zeta -1}) \\ 
-\frac{\zeta ^2}{4(2\ \zeta -1)^2b^2}([(2\ \zeta -1)^2-1]\vec R^{\prime
}-[(2\ \zeta -1)^2+1]\vec R)^2]+\frac{\alpha _2(m_1-m_2)}{%
4m_1m_2m_{12}b^2(2\ \zeta -1)}[-12\zeta (\zeta -1) \\ 
-\frac{2\ \zeta ^2}{b^2(2\ \zeta -1)}((\zeta -1)\vec R^{\prime }-\zeta \vec R%
)([(2\zeta -1)^2-1]\vec R^{\prime }-[(2\zeta -1)^2+1]\vec R)]+\frac 1{%
4m_1m_2b^2}[6\alpha _2(1+\frac{\alpha _2\zeta }{2\zeta -1}) \\ 
-\frac{4\zeta ^2\alpha _2^2}{b^2(2\ \zeta -1)^2}((2\zeta +1)\vec R^{\prime
}-(2\zeta -3)\vec R)^2]-i\frac{\beta _1\zeta ^2}{2m_{12}\alpha \ b^4(2\zeta
-1)}(1+\gamma )(\vec R\times \vec R^{\prime })\cdot (\frac{\stackrel{%
\rightharpoonup }{\sigma }_4}{m_1}-\frac{\stackrel{\rightharpoonup }{\sigma }%
_5}{m_2}) \\ 
+\frac 1{4m_1m_2}[(\frac{2\alpha _2\gamma }{b^2}+\frac{\gamma ^2}{b^2}\frac{%
2\zeta \alpha _2^2}{2\zeta -1})\vec \sigma _4\cdot \vec \sigma _5-\frac{%
4\gamma ^2\zeta ^2\alpha _2^2}{b^4(2\zeta -1)^2}(\zeta \vec R^{\prime
}-(\zeta -1)\vec R)\cdot \vec \sigma _4(\zeta \vec R^{\prime }-(\zeta -1)%
\vec R\cdot \vec \sigma _5]\} \\ 
-\frac{16\alpha _sC_{45}^t\zeta ^3}{\pi ^2b^6}(\frac{\beta _1}{2\zeta
-1-\zeta \alpha _2)})^{\frac 32}f_4^t(\vec R,\vec R^{\prime })_{ex}\frac 1{%
4m_1m_2}(\frac{m_1^2+m_2^2}{2m_1m_2}+\vec \sigma _4\cdot \vec \sigma _5)
\end{array}
\eqnum{B18}
\end{equation}
in which $\alpha _1=\frac{m_1}{m_1+m_2},\alpha _2=\frac{m_2}{m_1+m_2}$. The
other exchanged potential operators $\hat V_t^{ex}(\vec R,\vec R^{\prime
})^{24}$ and $\hat V_t^{ex}(\vec R,\vec R^{\prime })^{34}$ can directly be
written out from the above operators by changing the superscripts 14 to 24
and 34.

The potential operators corresponding to the terms of the direct part of the
potential in (B3) are shown in the following 
\begin{equation}
\begin{array}{c}
\hat V_{24}^D(\vec R,\vec R^{\prime })=\frac{16\pi \alpha _sC_{24}^d}b(\frac 
\zeta {\pi \ (2\zeta -\gamma )})^{\frac 32}\delta (\vec R-\vec R^{\prime
})e^{-\frac{\gamma \ \zeta }{b^2(2\zeta -\gamma )}\vec R^2}\{1-\frac 1{%
4m_1^2b^2}[3-\frac 34(\beta _2-\beta _1)^2(\frac 1{2\zeta }+\frac 1\gamma )
\\ 
-b^2((\frac 1{2\zeta }+\frac 1\gamma )\nabla _{\vec R}-(\frac 1{2\zeta }-%
\frac 1\gamma )\nabla _{\vec R^{\prime }})^2]+\frac 1{4m_1^2b^2}[3(1-\frac 1%
\gamma )-\frac{b^2}{\gamma ^2}(\nabla _{\vec R}-\nabla _{\vec R^{\prime
}})^2] \\ 
-i\frac 1{8m_1^2b^2}\frac{2(\beta _2-\beta _1)}{2\zeta -\lambda }(1+\gamma )(%
\vec \sigma _2-\vec \sigma _4)\cdot (\vec R\times \nabla _{\vec R})+i\frac 3{%
4m_1^2b^2(2\zeta -\lambda )}(1+\gamma )(\vec \sigma _2+\vec \sigma _4)\cdot (%
\vec R\times \nabla _{\vec R}) \\ 
+\frac 1{4m_1^2b^2}[\frac{2\zeta \ \lambda }{2\zeta -\lambda }\vec \sigma
_2\cdot \vec \sigma _4-\frac{4\gamma ^2\zeta ^2}{b^2(2\zeta -\gamma )^2}\vec 
R\cdot \vec \sigma _2\vec R\cdot \vec \sigma _4]\} \\ 
-\frac 1{4m_1^2}\frac{16\alpha _sC_{24}^d}{\pi ^2b^3}(\frac{\pi \ \zeta }{%
2\zeta -1})^{\frac 32}\delta (\vec R-\vec R^{\prime })e^{-\frac \zeta {%
b^2(2\zeta -1)}\vec R^2}\vec \sigma _2\cdot \vec \sigma _4.
\end{array}
\eqnum{B19}
\end{equation}
The potential operators corresponding to the terms $V_{14}^D(\vec R,\vec R%
^{\prime })$ and $V_{34}^D(\vec R,\vec R^{\prime })$ in (B3) have the same
form as shown above except for the subscripts 24 being changed to 14 and 34.
Introducing 
\begin{equation}
f_5^t(\vec R,\vec R^{\prime })_D=\exp \{-\frac 1{\zeta _1b^2(1-\frac{\zeta
_2^2}{\zeta _1^2})}(\vec R^2+2\frac{\zeta _2}{\zeta _1}\vec R\cdot \vec R%
^{\prime }+\vec R^{\prime 2})\}  \eqnum{B20}
\end{equation}
with $\zeta _1=1/(\zeta -\beta _1^2)+1/\gamma \alpha _2-2/\varsigma $ and $%
\zeta _2=1/(\zeta -\beta _1^2)+1/\gamma \alpha _2$ and 
\begin{equation}
f_6^t(\vec R,\vec R^{\prime })_D=\exp \{-\frac 1{\zeta _1^{\prime }b^2(1-%
\frac{\zeta _2^{\prime 2}}{\zeta _1^{\prime 2}})}(\vec R^2+2\frac{\zeta
_2^{\prime }}{\zeta _1^{\prime }}\vec R\cdot \vec R^{\prime }+\vec R^{\prime
2})\}  \eqnum{B21}
\end{equation}
with $\zeta _1^{\prime }=1/(\zeta -\beta _1^2)+1/\alpha _2-2/\varsigma $ and
and $\zeta _2^{\prime }=1/(\zeta -\beta _1^2)+1/\alpha _2$, the potential
operator corresponding to the term $V_{25}^D(\vec R,\vec R^{\prime })$ in
(B3) can be represented in the form 
\begin{equation}
\begin{array}{c}
\hat V_{25}^D(\vec R,\vec R^{\prime })=\frac{8\alpha _sC_{25}^d}{\pi
^2\alpha _2\zeta _1^3b^4}(\frac{\alpha \ \zeta }{\gamma (\zeta ^2-\beta
_1^2)(1-\frac{\zeta _2^2}{\zeta _1^2})})^{\frac 32}f_5^t(\vec R,\vec R%
^{\prime })_D\{1-\frac 1{m_{12}^2b^2}[\frac 3{2\alpha _1}+\frac{(\alpha
_1-\beta _1)^2}{4\alpha _1^2}(-\frac 6{\zeta -\beta _1^2}+ \\ 
\frac{12}{(\zeta -\beta _1^2)^2\zeta _1(1+\frac{\zeta _2}{\zeta _1})})-\frac{%
(\alpha _1-\beta _1)^2}{4\alpha _1^2b^2}\frac 4{(\zeta -\beta _1^2)^2(\zeta
_1+\zeta _2)^2}(\vec R-\vec R^{\prime })^2]+\frac{m_1-m_2}{4m_1m_2m_{12}b^4}%
\alpha _2(\beta _2-\frac{\beta _1}\alpha )[(-\frac 6{\zeta -\beta _1^2} \\ 
+\frac{12}{(\zeta -\beta _1^2)^2\zeta _1(1+\frac{\zeta _2}{\zeta _1})})b^2-%
\frac 4{(\zeta -\beta _1^2)^2(\zeta _1+\zeta _2)^2}(\vec R-\vec R^{\prime
})^2)]+\frac 1{4m_1m_2}[\frac{6\alpha _2}{b^2}+\frac{\alpha _2^2}{b^4}((-%
\frac 6{\zeta -\beta _1^2}+\frac{12}{(\zeta -\beta _1^2)^2\zeta _1(1+\frac{%
\zeta _2}{\zeta _1})})b^2 \\ 
-\frac 4{(\zeta -\beta _1^2)^2(\zeta _1+\zeta _2)^2}(\vec R-\vec R^{\prime
})^2)]-i\frac{4\ (\alpha _1-\beta _1)}{4m_{12}\alpha _1\ (\zeta -\beta
_1^2)b^4(\zeta _1^2-\zeta _2^2)}(1+\gamma )(\vec R\times \vec R^{\prime
})\cdot (\frac{\vec \sigma _2}{m_1}-\frac{\vec \sigma _5}{m_2}) \\ 
+i\frac{4\alpha _2}{4m_1m_2\ (\zeta -\beta _1^2)b^4(\zeta _1^2-\zeta _2^2)}%
(1+\gamma )(\vec R\times \vec R^{\prime })\cdot [(2+\frac{m_2}{m_1})\vec 
\sigma _2-(2+\frac{m_1}{m_2})\vec \sigma _5] \\ 
+\frac 1{4m_1m_2}[\frac 4{b^2(\zeta _1-\zeta _2)}\vec \sigma _2\cdot \vec 
\sigma _5-\frac 4{b^4(\zeta _1-\zeta _2)^2}(\vec R+\vec R^{\prime })\cdot 
\vec \sigma _2(\vec R+\vec R^{\prime })\cdot \vec \sigma _5]\} \\ 
-\frac{8\alpha _sC_{25}^d}{\pi ^2\alpha _2\zeta _1^{\prime 3}b^4}(\frac{%
\alpha \ \zeta }{(\zeta ^2-\beta _1^2)(1-\frac{\zeta _2^{\prime 2}}{\zeta
_1^{\prime 2}})})^{\frac 32}f_6^t(\vec R,\vec R^{\prime })_D\frac 1{4m_1m_2}(%
\frac{m_1^2+m_2^2}{2m_1^2m_2^2}+\vec \sigma _2\cdot \vec \sigma _5).
\end{array}
\eqnum{B22}
\end{equation}
The potential operators corresponding to the remaining two terms in (B3) are
of the same form as shown above except for the subscripts 25 being replaced
by 15 and 35.

Let us turn to the $\overline{K}N$ interaction potential. For the $\overline{%
K}N$ interaction, the effective potential coming from the $t$-channel OGEP
only has a direct part as represented in (B3) and (B19)-(B22) because there
are no identical particles between the $N$-cluster $(qqq)$ and the $%
\overline{K}$-cluster $(\overline{q}s)$. Besides, the nonlocal effective
potential derived from the $s$-channel OGEP must be considered and plays an
essential role in the $\overline{K}N$ interaction. This potential can be
written as

\begin{equation}
V^s(\vec R,\vec R^{\prime })=V_{14}^{s\_d}(\vec R,\vec R^{\prime
})+V_{24}^{s\_d}(\vec R,\vec R^{\prime })+V_{34}^{s\_d}(\vec R,\vec R%
^{\prime })  \eqnum{B23}
\end{equation}
where $V_{ij}^{s\_d}(\vec R,\vec R^{\prime })$ denotes the direct term of
the potential generated from the interaction between the quark $i$ and the
antiquark $j$. The corresponding operator form of the potentials on the RHS
of (B23) can be derived in the same way as shown before for the $t$-channel
potential operators. The results are 
\begin{equation}
\begin{array}{c}
\hat V_{14}^{s\_d}(\vec R,\vec R^{\prime })=\frac{\pi \alpha
_sF_{14}^aC_{14}^s}{2m_1^2(2\pi b^2)^{3/2}}(\frac{2\zeta }{2\zeta -1})^{%
\frac 32}\delta (\vec R-\vec R^{\prime })e^{-\frac \zeta {4\ b^2(2\zeta -1)}(%
\vec R+\vec R^{\prime })^2}\{(3+\vec \sigma _1\cdot \vec \sigma _4) \\ 
-\frac 1{4m_1^2b^2}[3-\frac{3(\beta _2-\beta _1)^2}{2\zeta }+\frac{3\ (\beta
_2-\beta _1)^2}{\zeta \ (2\zeta -1)}-\frac{2\ (\beta _2-\beta _1)^2}{\zeta
(2\zeta -1)^2b^2}\vec R^2+\frac{4\ (\beta _2-\beta _1)^2}{(2\zeta -1)\zeta }%
\vec R\cdot \nabla _{\vec R}+b^2\nabla _{\vec R}^2]\cdot (3+\vec \sigma
_1\cdot \vec \sigma _4) \\ 
-\frac 1{2m_1^2b^2}[3-6(\frac 1{4\zeta }-\frac \zeta {2\zeta -1})+\frac{%
\zeta ^2}{b^2\ (2\zeta -1)^2}(\vec R+\vec R^{\prime })^2+\frac 3{\zeta \
(2\zeta -1)}-\frac 2{(2\zeta -1)^2b^2}\vec R^2 \\ 
+\frac 4{(2\zeta -1)\zeta }\vec R\cdot \nabla _{\vec R}-\frac{2b^2}{\zeta ^2}%
\nabla _{\vec R}^2]\cdot (2+\vec \sigma _1\cdot \vec \sigma _4)-i\frac{\beta
_2-\beta _1}{m_1^2b^2(2\zeta -1)}(\vec \sigma _1-\vec \sigma _4)\cdot (\vec R%
\times \nabla _{\vec R}) \\ 
+\frac 1{2m_1^2b^2}[-\frac{2\zeta }{2\zeta -1}\vec \sigma _1\cdot \vec \sigma
_4+\frac 1{4b^2(2\zeta -1)^2}(\vec R+\vec R^{\prime })\cdot \vec \sigma _1(%
\vec R+\vec R^{\prime })\cdot \vec \sigma _4+\frac{4\zeta }{2\zeta -1}\vec 
\sigma _1\cdot \vec \sigma _4 \\ 
-\frac{8\zeta ^2}{(2\zeta -1)^2b^2}\vec R\cdot \vec \sigma _1\vec R\cdot 
\vec \sigma _4+\frac{16\zeta }{2\zeta -1}\vec R\cdot \vec \sigma _1\nabla _{%
\vec R}\cdot \vec \sigma _4+8b^2\ \nabla _{\vec R}\cdot \vec \sigma _1\nabla
_{\vec R}\cdot \vec \sigma _4]\}.
\end{array}
\eqnum{B24}
\end{equation}
The other two terms $\hat V_{24}^{s\_d}(\vec R,\vec R^{\prime })$ and $\hat V%
_{34}^{s\_d}(\vec R,\vec R^{\prime })$ can be written out from the above
expression by the substitution of the subscripts 24 and 34 for 14.

The effective $KN$ ($\overline{K}N$) potential derived from the interquark
harmonic oscillator confining potential and quark interchanges is of a
simple expression whose operator form is represented as 
\begin{equation}
\begin{array}{c}
\hat V_c^{ex}(\vec R,\vec R^{\prime })=-12b^2\omega ^2\{C_s^{24}\mu
_{24}+C_s^{34}\mu _{34}+C_s^{12}\mu _{12}+C_s^{13}\mu _{13}+C_s^{14}\mu _{14}
\\ 
+\frac 1{2\alpha _2}[C_s^{25}\mu _{25}+C_s^{35}\mu _{35}+C_s^{15}\mu
_{15}+C_s^{45}\mu _{45}]\}4(\frac \zeta \pi )^3(\frac{\pi \zeta }{(2\zeta
-1)b^2})^{\frac 32}e^{^{-\frac{\zeta (2\zeta -1)}{4b^2}(\vec R-\vec R%
^{\prime })^2-\frac \zeta {4b^2(2\zeta -1)}(\vec R+\vec R^{\prime })^2}}
\end{array}
\eqnum{B25}
\end{equation}
where $\mu _{ij}$ denotes the reduced mass of interacting quarks $i$ and $j$
and the color factors are the same as the ones appearing in the $t$-channel
effective potentials.

Apart from the potentials listed above, there are additional terms in the $%
KN $ and $\overline{K}N$ potentials occurring in the resonating group
equation which arise from the kinetic term and the normalization term in the
resonating group equation due to the effect of quark rearrangement. They are
respectively written in the following: 
\begin{equation}
\begin{array}{c}
T^{ex}(\vec R,\vec R^{\prime })=(-3)\alpha ^{3/2}\{\frac 1{2\mu }[\frac \zeta
{2b^2}-\frac{(\zeta \,\vec R-\zeta \,\vec R^{\prime }+\vec R^{\prime })^2}{%
4b^4}]+\frac 1{2\mu _1}[\frac 3{4b^2}-\frac{\vec R^{\prime 2}}{16b^4}] \\ 
+\frac 1{2\mu _2}[\frac 1{4b^2}-\frac{\vec R^{\prime 2}}{144b^4}]+\frac 1{%
2\mu _3}[\frac{3\alpha _2}{2\text{ }b^2}-\frac{\alpha _2^2\vec R^{\prime 2}}{%
4b^4}]\}f_T^{ex}(\vec R,\vec R^{\prime })
\end{array}
\eqnum{B26}
\end{equation}
where 
\begin{equation}
\begin{array}{c}
f_T^{ex}(\vec R,\vec R^{\prime })=4\zeta ^3(\frac \zeta {\pi b^2(2\zeta
-1)(2\zeta -\frac 2{\frac 13+\alpha _1}+1)})^{\frac 32} \\ 
\times \exp \{-\frac{\zeta (\frac 2{\frac 13+\alpha _1}-1)}{4b^2(2\zeta -%
\frac 2{\frac 13+\alpha _1}+1)}(\vec R-\vec R^{\prime })^2-\frac \zeta {%
4b^2(2\zeta -1)}(\vec R+\vec R^{\prime })^2\},
\end{array}
\eqnum{B27}
\end{equation}
with 
\begin{equation}
\mu =\frac{3m_1(m_1+m_2)}{4m_1+m_2},\mu _1=\frac{m_1}2,\mu _2=\frac{2m_1}3%
,\mu _3=\frac{m_1m_2}{m_1+m_2},\gamma _1=3/(1/3+\alpha _1).  \eqnum{B28}
\end{equation}
and

\begin{equation}
N^{ex}(\vec R,\vec R^{\prime })=-3E_r^{\frac 32}\exp \{-\frac{\zeta (-2\beta
_1^2-2\beta _2^2-1)}{4b^2(2\zeta -2\beta _1^2-2\beta _2^2+1)}(\vec R-\vec R%
^{\prime })^2-\frac \zeta {4b^2(2\zeta -1)}(\vec R+\vec R^{\prime })^2\} 
\eqnum{B29}
\end{equation}
here $E_r$ is the relative energy of two clusters.

The effective potential in Eq. (23) is given by the color-spin-isospin
matrix element of the above potential operator 
\begin{equation}
V(\vec R,\vec R^{\prime })=\langle \Psi _{TM\frac 12m}(1,2,3,4,5)\mid \hat V(%
\vec R,\vec R^{\prime })\mid \Psi _{TM\frac 12m}(1,2,3,4,5)  \eqnum{B30}
\end{equation}
where $\hat V(\vec R,\vec R^{\prime })$ has the expression as written in Eq.
(24) and in this appendix and $\Psi _{TM\frac 12m}(1,2,3,4,5)$ is the
color-spin-isospin wave function represented in Eq. (12) and Appendix A.

To incorporate the QCD renormalization effect into the model, the QCD fine
structure constant $\alpha _s$ and quark masses in the potential operators
will be replaced by their effective ones. The effective fine structure
constant derived in the one-loop approximation has the expression like this
[40] 
\begin{equation}
\alpha _s(\lambda )=\frac{\alpha _s^0}{1+\frac{\alpha _s^0}{2\pi }G(\lambda )%
}  \eqnum{B31}
\end{equation}
where $\alpha _s^o$ is a coupling constant and $G(\lambda )$ is a function
of variable $\lambda $ which has different expressions given by the
time-like momentum subtraction (the subtraction performed at time-like
renormalization point) and the space-like momentum subtraction (the
subtraction carried out at the space-like renormalization point). For the
time-like momentum subtraction, 
\begin{equation}
G(\lambda )=11\ln \lambda -\frac 23N_f[2+\sqrt{3}\pi -\frac 2{\lambda ^2}+(%
\frac 2{\lambda ^2}+1)\frac{\sqrt{\lambda ^2-4}}\lambda \ln \frac 12(\lambda
+\sqrt{\lambda ^2-4})]  \eqnum{B32}
\end{equation}
where $N_f$ is the quark flavor number which will be taken to be three in
this paper. While, for the space-like momentum subtraction, 
\begin{equation}
\begin{tabular}{l}
$G(\lambda )=11\ln \lambda -\frac 23N_f[\frac 2{\lambda ^2}-2-(\frac 2{%
\lambda ^2}-1)\frac{\sqrt{\lambda ^2+4}}\lambda \ln \frac 12(\lambda +\sqrt{%
\lambda ^2+4})$ \\ 
$+\sqrt{5}\ln \frac 12(1+\sqrt{5})]$%
\end{tabular}
\eqnum{B33}
\end{equation}
in which $\lambda $ is defined as $\lambda =\sqrt{q^2/\mu ^2}$ with $q$
being a momentum variable and $\mu $ the fixed scale parameter. It is noted
that in writing the above effective coupling constant, the mass difference
between different quarks is ignored for simplicity. In this paper, the
effective coupling constants given in the space-like momentum subtraction
and the time-like momentum subtraction are suitable for the $t$- channel
OGEP and the $s$-channel OGEP, respectively.

The effective quark mass is represented as 
\begin{equation}
m_R(\lambda )=m_Re^{-S(\lambda )}  \eqnum{B34}
\end{equation}
where $m_R$ is the constant quark mass given at $\lambda =1$ which will
appropriately be chosen to be the constituent quark mass in the quark
potential model, $S(\lambda )$ is a function which also has different
expressions for the different subtractions. For the time-like momentum
subtraction, 
\begin{equation}
S(\lambda )=\frac{\alpha _R^0}\pi \frac{1-\lambda }\lambda \{2+(\frac 2{%
\lambda ^2}-\frac{1+\lambda }{\lambda ^2})\ln \mid 1-\lambda ^2\mid \}. 
\eqnum{B35}
\end{equation}
In this paper, only the effective quark masses given in the time-like
momentum subtraction is necessary to be considered.

\section{Figure Captions}

Fig.1: The theoretical $\overline{K}N$ S-wave phase shifts in the $I=0$ and $%
1$ channels.

Fig.2: The theoretical $\overline{K}N$ $P_{01}$-wave and $P_{13}$-wave phase
shifts.

Fig.3: The theoretical $KN$ $P_{01}$ phase shift. The experimental phase
shift [41, 42] are shown by black squares with error bars.

Fig.4: The theoretical $KN$ $D_{03}$ phase shift. The experimental phase
shift [41, 42] are shown by black squares with error bars.

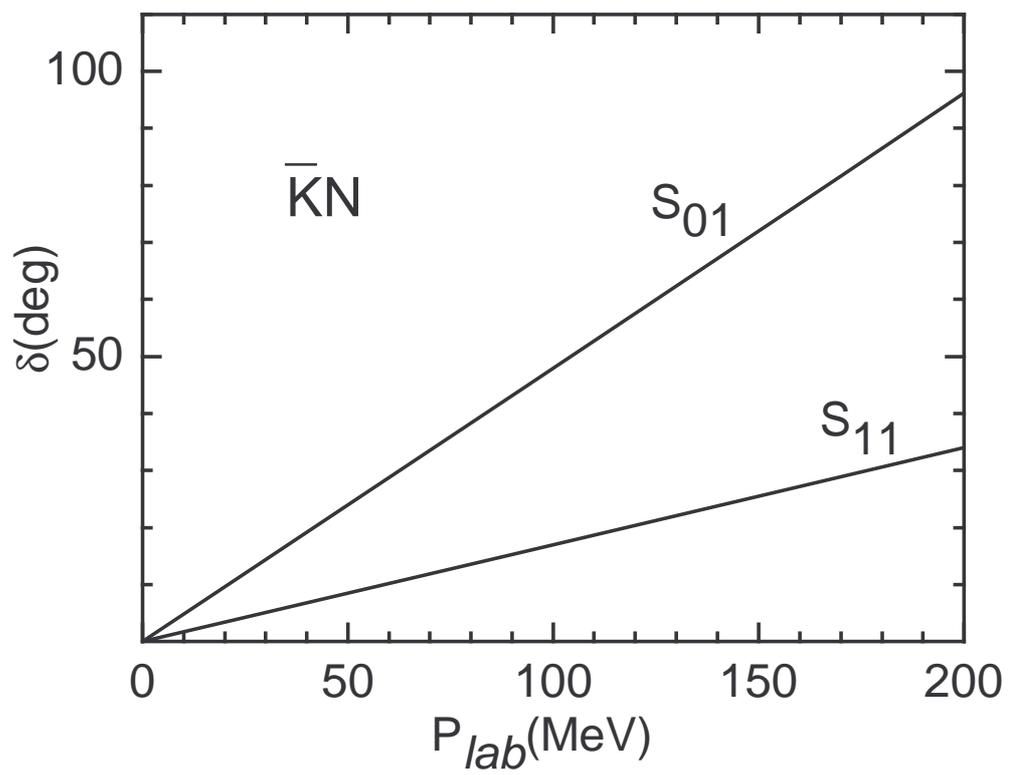
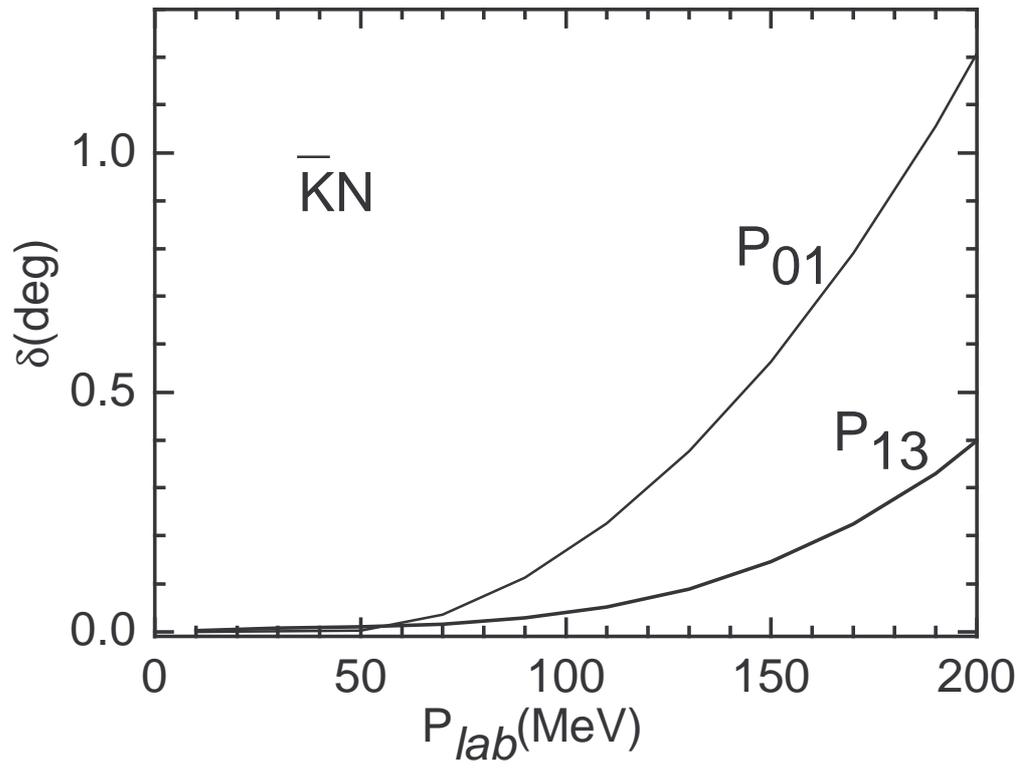
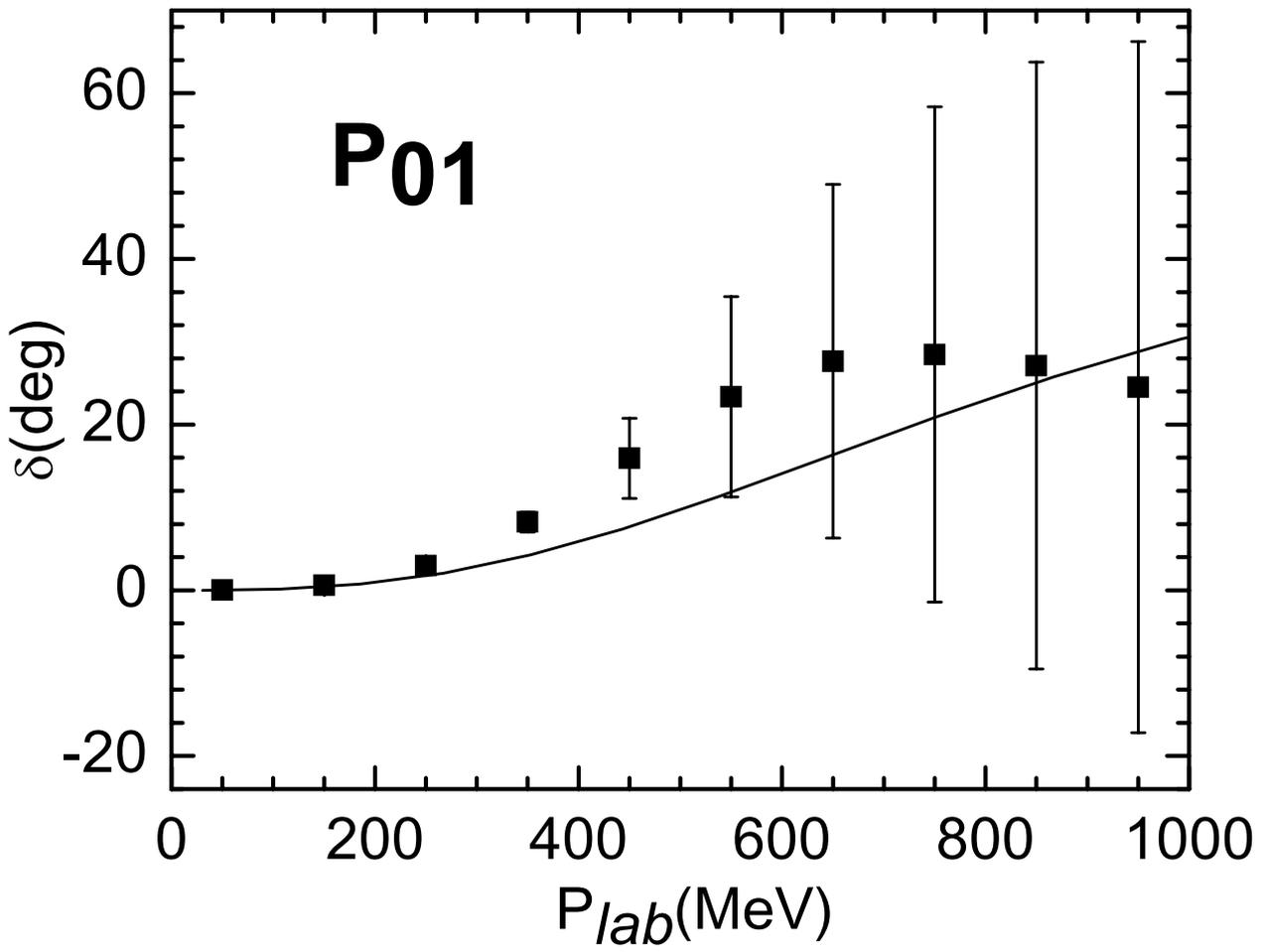
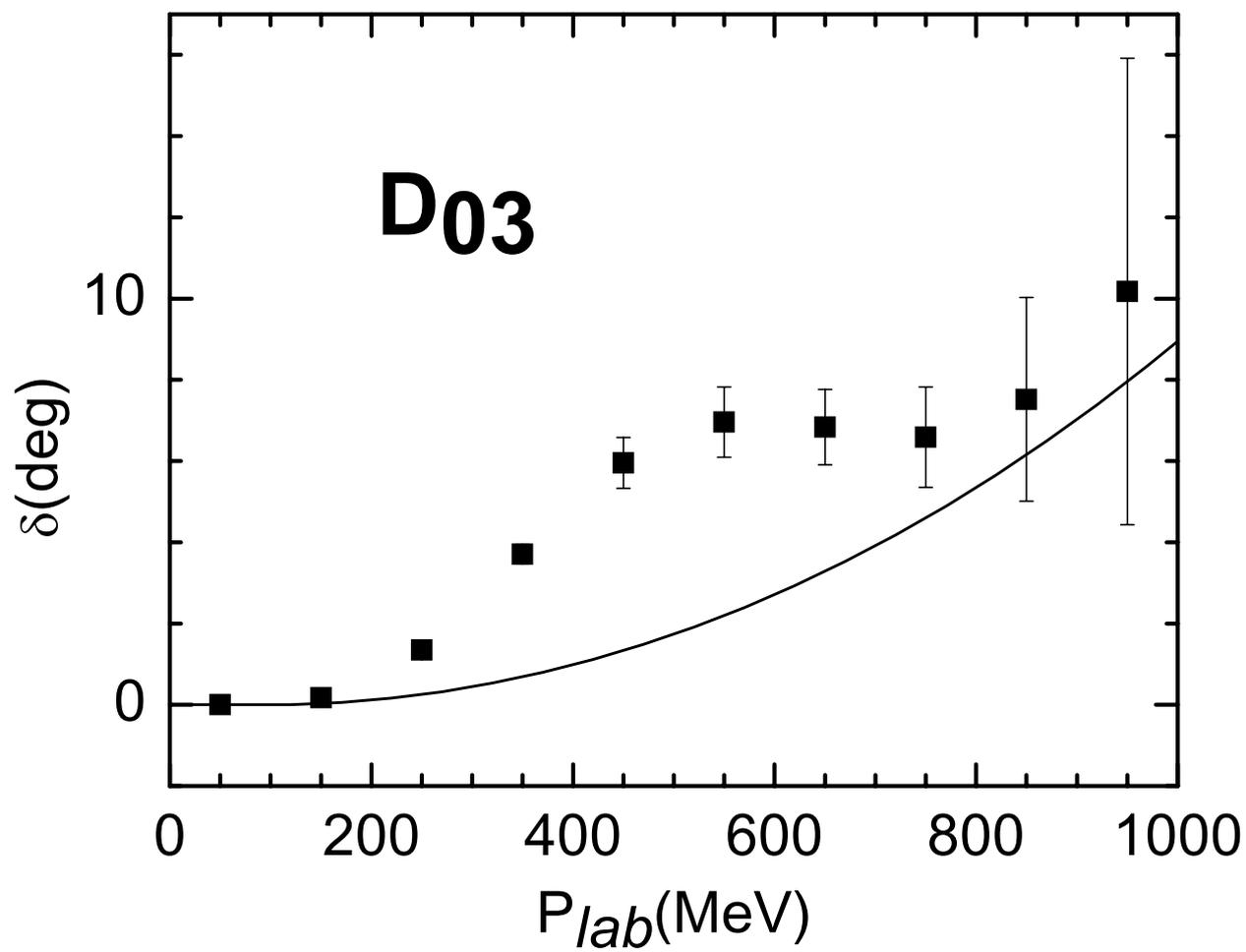
\end{document}